\documentclass[aps,pra,twocolumn,superscriptaddress,longbibliography,footinbib]{revtex4-2}  %
\usepackage{graphicx} 
\usepackage{float} 
\usepackage{dcolumn} 
\usepackage{bm,color}
\usepackage{amsmath,amssymb,dsfont,amstext,amsfonts}
\usepackage{extarrows}
\usepackage[colorlinks=true,linkcolor=blue,urlcolor=blue,citecolor=blue]{hyperref}
\usepackage{xcolor}
\usepackage{wasysym}
\usepackage{mathtools}
\usepackage{bbold}
\usepackage[export]{adjustbox}
\usepackage{mathdots}
\usepackage[normalem]{ulem} 
\usepackage{color}

\def\bra#1{\mathinner{\langle{#1}|}}
\def\ket#1{\mathinner{|{#1}\rangle}}

\def\braket#1{\mathinner{\langle{#1}\rangle}}

\def\sgn{\mathrm{sgn}}
\def\re{\mathrm{Re}\,}
\def\im{\mathrm{Im}\,}
\def\tr{\mathrm{Tr}}

\newcommand{\sect}[1]{\vspace{0.3em}{\it \textcolor{blue}{#1.---}}}

\begin{document}
\title{Rational Approximations of Quasi-Periodic Problems via Projected Green's Functions}
\author{Dan S. Borgnia}
\email{dbognia@g.harvard.edu}
\affiliation{Department of Physics, Harvard University, Cambridge, MA 02138}
\author{Ashvin Vishwanath}
\affiliation{Department of Physics, Harvard University, Cambridge, MA 02138}
\author{Robert-Jan Slager}
\email{rjs269@cam.ac.uk}
\affiliation{TCM Group, Cavendish Laboratory, University of Cambridge, J. J. Thomson Avenue, Cambridge CB3 0HE, United Kingdom}
\affiliation{Department of Physics, Harvard University, Cambridge, MA 02138}
\date{\today}

\begin{abstract}
We introduce the projected Green's function technique to study quasi-periodic systems such as the Andre-Aubry-Harper (AAH) model and beyond. In particular, we use projected Green's functions to construct a ``rational approximate" sequence of transfer matrix equations consistent with quasi-periodic topology, where convergence of these sequences corresponds to the existence of extended eigenfunctions. We motivate this framework by applying it to a few well studied cases such as the almost-Mathieu operator (AAH model), as well as more generic non-dual models that challenge standard routines. The technique is flexible and can be used to extract both analytic and numerical results, e.g. we analytically extract a modified phase diagram for Liouville irrationals. As a numerical tool, it does not require the fixing of boundary conditions and circumvents a primary failing of numerical techniques in quasi-periodic systems, extrapolation from finite size. Instead, it uses finite size scaling to define convergence bounds on the full irrational limit.
\end{abstract}
\maketitle

\sect{Introduction} Quasi-periodic systems have been heavily studied in recent years in many contexts, i.e. twisted bi-layer systems, incommensurate Floquet Hamiltonians, localization transitions in 1D, and virtural topological invariants \cite{prodan2015,kraus2012topological,jitomirskaya2019critical,PhysRevB.101.014205, tbg2, spirals, refealhusemblquasi, slonghiphasetrannonher,tbg1}. 
New quasi-periodic models have been proposed in each of these contexts, prompting many numerical pursuits in search of metal insulator transitions, novel topological phases, and 1D mobility edges \cite{PhysRevB.101.014205,slonghiphasetrannonher,refealhusemblquasi, Wang_2021qp, modak,ganeshan2015nearest}. However, quasi-periodic systems are notoriously difficult to simulate, with many numerical \cite{PhysRevLett.109.116404,jitomirskayatalk} and even some analytic results \cite{madsen2013topological,prodan2015} failing to capture the full behavior of these regular aperiodic systems. Beyond the fundamental challenges of exploring phase transitions, quasi-periodic systems are impossible to exactly simulate due to the infinite period of a quasi-periodic potential, and approximations are very sensitive to boundary conditions -- in-gap edge modes can arise \cite{prodan2015}. This letter presents an alternative approach resting on the approximation of quasi-periodic transfer matrices via a continued fraction sequence of higher dimensional rational projected Green's functions. Our approach both provides a measure of numerical accuracy (explicit finite size convergence conditions) and circumvents the definition of boundary conditions by resting on projected Green's function (pGf) technology, a tool from translation-invariant systems \cite{Slager2015,slager2019translational,Borgnia2020,mong2011edge,volovik2003universe,Gurarie2011} defined by the bulk Green's function and carrying boundary information.

The paradigmatic quasi-periodic system is the 1D Andre-Aubry-Harper (AAH) Model (almost-Mathieu operator in the math community) \cite{aubry1980annals,aubry1981bifurcation,frohlich1990localization,bellissard1982quasiperiodic,bellissard1986gaplabeling,bellissard1986k,avila2006reducibility,avila2006solving,jitomirskaya1998anderson,jitomirskaya1999metal,jitomirskaya2012analytic,jitomirskaya2019critical,avila2017sharp}. Famous for its sharp 1D metal insulator transition (MIT), self-duality, and mapping to a 2D Hofstadter Hamiltonian \cite{aubry1980annals,aubry1981bifurcation,frohlich1990localization,bellissard1982quasiperiodic,bellissard1986gaplabeling,bellissard1986k,avila2006reducibility,avila2006solving,jitomirskaya1998anderson,jitomirskaya1999metal,jitomirskaya2012analytic,jitomirskaya2019critical,avila2017sharp}, the AAH model is an important benchmark for any tool -- many current methods fail to find a complete phase diagram \cite{jitomirskaya1998anderson,jitomirskaya1999metal,jitomirskaya2012analytic,jitomirskaya2019critical,avila2017sharp}. We therefore construct our formalism in relation to the AAH model and then address generalizations to other 1D quasi-periodic models, self-dual and beyond. Higher dimensions follow naturally, but are left to future work. This letter proceeds by introducing rational approximates to quasi-periodic operators, constructing a transfer matrix equation (TME) for each rational approximate, detailing a convergence criteria for these rational TME approximates, extracting the spatial properties of quasi-periodic eigenfunctions from the aforementioned criteria, and finally, showcasing a few generalizations beyond the AAH model afforded by our approach.

\begin{figure*}
    \centering
    \includegraphics[scale=.45]{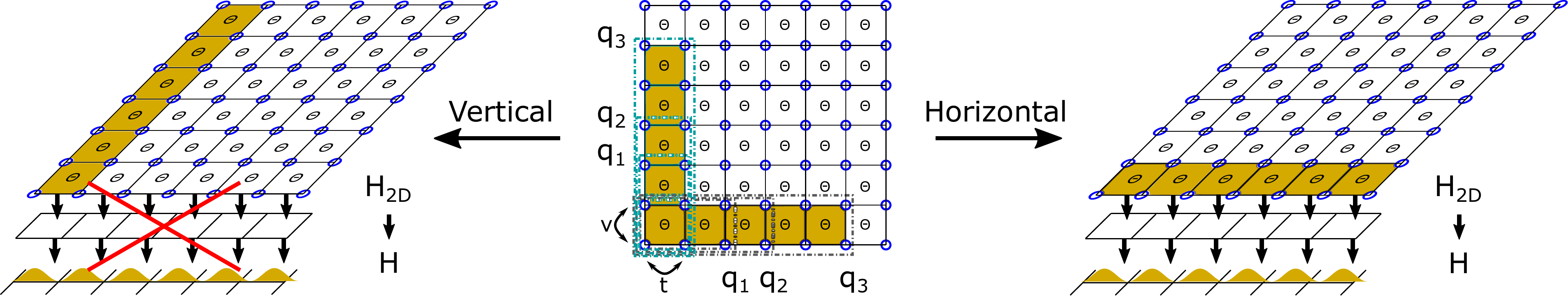}
    \caption{Magnetic unit cells are chosen in the 2D parent Hamiltonian (center), but not all Magnetic unit cells have a convergent rational approximate sequence, see S.I.~\ref{siboundstransfer}. The convergent alignment defines allowed projections back to 1D (left/right), i.e. horizontal unit cells naturally project back to 1D, while vertical unit cells do not. Concretely, for the AAH model, Eq. \eqref{2Dhamiltoniangauge} the regimes $V<t$ and $V>t$ dictate a horizontal or vertical cell respectively. This is rooted in the topology of the 2D parent system that defines the projection to the 1D quasi-periodic model. }
    \label{fig1}
\end{figure*}
\sect{Rational Approximates}
Quasi-periodic  systems can be readily analyzed with sequences of \textit{rational approximates}. These are linear operators whose quasi-periodic  parameter, $\alpha\notin\mathbb{Q}$ is replaced by a rational parameter $\frac{p_n}{q_n}\in\mathbb{Q}$ such that $\lim_{n\rightarrow\infty}\frac{p_n}{q_n} = \alpha$. The fastest converging sequence is the continued fraction approximation for $\alpha$, i.e. $\sqrt{2}-1 = \frac{1}{2+\frac{1}{2+\frac{1}{\ldots}}}$. We can choose 
\[\frac{p_n}{q_n} = a_0 + \cfrac{1}{a_1 +  \cfrac{1}{ \ddots + \cfrac{1}{a_n} }}.\]

Different irrational numbers have different rational approximates whose convergence properties can be vastly different. For example, the golden mean and $\sqrt{2}$ are Diophantine numbers and their continued fraction approximation converges with at worst $\vert\alpha - \frac{p_n}{q_n}\vert<\frac{1}{\sqrt{5}q_n^{2}}$ -- saturated by the golden mean. On the other hand, Liouville numbers, discussed below, are exponentially well approximated by their continued fractions, e.g. Lioville's constant defined $L\equiv \sum_{n=1}^{\infty} 10^{-n!}$.

The main results of this work generalize to multiple classes of quasi-periodic  models, but since the Andre-Aubry model 
is the most well studied we will use it as a test bed and subsequently generalize our findings. 
Hence, we recall the simple AAH Hamiltonian as given by
\begin{align}\label{HamiltonianAAHeqRS}
\hat{H} = \sum_{x} t(\hat{c}^{\dagger}_{x+1}\hat{c}_{x}+\hat{c}_{x+1}\hat{c}^{\dagger}_{x})+2V\cos(\Theta x+\delta)\hat{c}^{\dagger}_{x}\hat{c}_{x},
\end{align}
with $\Theta = 2\pi \alpha$ and $\alpha\in\mathbb{R}-\mathbb{Q}$. Note, the case where $\alpha\in\mathbb{Q}$ is called the Mathieu operator and is simply a periodic 1D band model. One can take a sequence of Mathieu operators 
\begin{align}
\hat{H}_{N} = \sum_{x} t(\hat{c}^{\dagger}_{x+1}\hat{c}_{x}+\hat{c}_{x+1}\hat{c}^{\dagger}_{x})+2V\cos(\Theta_{N} x)\hat{c}^{\dagger}_{x}\hat{c}_{x}
\end{align}
with $\Theta_{N} = 2\pi \frac{p_N}{q_N}$ as the rational approximate sequence for the almost-Mathieu operator, where we have suppressed the phase $\delta$. Unfortunately, we must choose our approximates more carefully. The AAH model and numerous other 1D quasi-periodic systems have non-trivial bulk topology arising from phase shift degrees of freedom in the quasi-periodic pattern, i.e. $\delta \rightarrow\delta + \Theta$ in the potential is equivalent to the translation $x\rightarrow x+1$ on the lattice \cite{prodan2015,bellissard1982quasiperiodic,bellissard1986gaplabeling}.  As discussed in supplemental information (S.I.)~\ref{siaahalgebra}, the spectral gaps of quasiperiodic systems as a function of the incommensurate parameter, $\alpha \in [0,1]$, can be labeled by integers, Fig.~\ref{patternfigure}a, roughly corresponding to the Chern number of the gap. 

Stacking de-coupled AAH chains and parameterizing each layer by the phase choice $\delta_{y}$ \cite{jitomirskaya1998anderson,prodan2015,bellissard1982quasiperiodic,kraus2012topological} provides an intuitive picture of the underlying bulk topology. Naively, the phase choice is irrelevant as it corresponds to a shift in initial position of an infinite chain, but the resulting 2D {\it parent} Hamiltonian, as function of $x$ and $\delta_{y}$, has a topological notion. In particular, by taking an inverse Fourier transform along the $\delta_{y}$ coordinate, it corresponds to a 2D tight-binding model with an irrational magnetic flux per plaquette, see S.I.~\ref{siaahalgebra}. 
\begin{eqnarray}\label{2Dhamiltoniangauge}
    \mathcal{H}_{2D} &=&\sum_{x,\delta_{y}}t \hat{c}_{x+1,\delta_{y}}^{\dagger}\hat{c}_{x,\delta_{y}} + t^{*} \hat{c}_{x,\delta_{y}}^{\dagger}\hat{c}_{x+1,\delta_{y}}\nonumber\\
    &+& 2V\cos(\Theta x+\delta_{y})\hat{c}_{x,\delta_{y}}^{\dagger}\hat{c}_{x,\delta_{y}},\\\label{2Dhamiltonian}
    \tilde{\mathcal{H}}_{2D}  &=&\sum_{x,y}t (\hat{c}_{x+1,y}^{\dagger}\hat{c}_{x,y} + \hat{c}_{x,y}^{\dagger}\hat{c}_{x+1,y} )\nonumber\\
    &+& V(e^{i\Theta x}\hat{c}_{x,y+1}^{\dagger}\hat{c}_{x,y} + e^{-i\Theta x}\hat{c}_{x,y-1}^{\dagger}\hat{c}_{x,y}).
\end{eqnarray}
The 2D spectrum amounts to a Hofstadter butterfly when varying the flux per plaquette, $\Theta$. For any rational flux, $\Theta/2\pi = p/q \in\mathbb{Q}$, one can define a magnetic unit cell specifying bands that have a Chern number, which sum to zero. This is however not possible for an irrational flux. Since the the rational sequence of transfer matrices is sensitive to pGF zeros, whose existence indicate non trivial bulk topology as detailed in the subsequent, one needs to require any approximating sequence to have equivalent bulk topology in the limit $N\rightarrow\infty$ to the full quasi-periodic Hamiltonian.

One way to respect the non-trival quasi-periodic topology is to choose our rational approximates in the 2D parent Hamiltonian description. As the Hamiltonian in Eq.~\eqref{2Dhamiltonian} defines the gap labeling sequence of the fully incommensurate flux \cite{PhysRevB.91.014108,bellissard1982quasiperiodic,bellissard1986gaplabeling}. We can therefore create a sequence of rational approximates to an irrational flux $\lbrace a_{n}\rbrace$, such that $\lim_{n\rightarrow\infty} \frac{p_{n}}{q_n} = \alpha$.

\sect{Transfer Matrix Theory} For each rational approximate, $H_{N}$, we construct a transfer matrix equation (TME) -- see S.I.~\ref{transfermatrixsi} and Ref. \cite{dwivedi2016bulk} for more details. The unit cell for the N-th rational approximate is $q_{N}$ sites long, and a wavefunction on the n-th unit cell is defined by
\begin{eqnarray}
\Psi_{n,N} = \begin{pmatrix}\psi_{n+1}& \ldots & \psi_{n+q_{N}}\end{pmatrix}^{T}.
\end{eqnarray}
This translates the eigenvalue equation into a simple form,
\begin{eqnarray}\label{basictme}
J_{N}\Psi_{n+1}+M_{N}\Psi_{n}+J_{N}^{\dagger}\Psi_{n-1} = E \Psi_{n}.
\end{eqnarray}
Here $J_{N}$ is the hopping matrix connecting the $q_{N}$-th site of $n$-th unit cell to the $1$-st site of the $n+1$-th unit cell, $J_{N}^{\dagger}$ does the opposite, and $M_{N}$ is the intra-unit cell term which acts internally on the $q_{N}$ internal sites of $\Psi_{n}$, see S.I.~\ref{transfermatrixsi}. 

From this we construct a TME,
\begin{eqnarray}\label{transfermatrixeq}
\begin{pmatrix}
J_{N}^{-1}(E-M_{N}) & -J_{N}^{-1}J_{N}^{\dagger}\\ 1  & 0
\end{pmatrix}\begin{pmatrix}
\Psi_{n} \\ \Psi_{n-1}
\end{pmatrix} = \begin{pmatrix}
\Psi_{n+1} \\ \Psi_{n}
\end{pmatrix}. 
\end{eqnarray}
Note, the assumed invertibility of the nearest neighbor (rank 1) hopping matrix, $J_N$. For larger unit cells, $J_N$ will in general not be invertible. Taking advantage of the detailed work in \cite{dwivedi2016bulk}, we use that $\text{rank}(J_{N}) = 1$ to reduce the transfer matrix in Eq~\eqref{transfermatrixeq}, from a $2q_{N}\times 2q_{N}$ matrix to a $2\times2$ matrix for each rational approximate, see S.I.~\ref{transfermatrixsi}.

\begin{figure}
    \centering
    \includegraphics[scale=0.32]{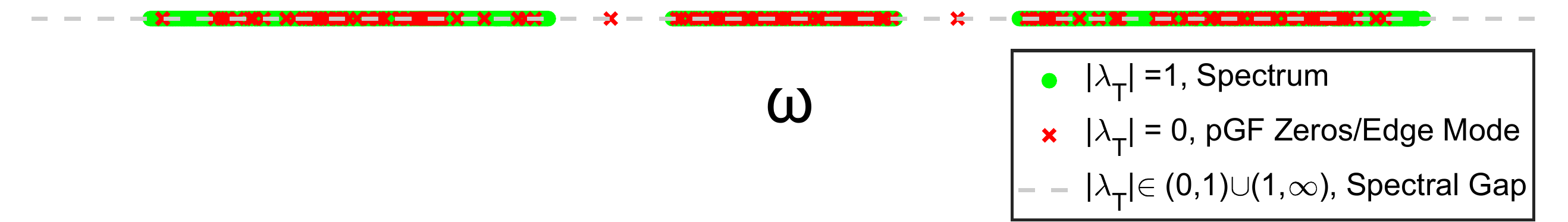}
    \caption{ Numerically computed AAH spectrum (green), corresponding spectral gaps (grey), and pGF zeros (red x) for parameters $t =1, V = 0.5, N = 2048, \alpha = \frac{1}{2}(\sqrt{5}-1)$. Each point is labeled by its transfer matrix eigenvalues. Spectrum (green) has transfer matrix eigenvalues on the unit circle, $\vert\lambda_{T}\vert=1$. Zero eigenvalues of the pGF (red x) imply the transfer matrix is rank deficient, $\lambda_T=0$. Note the many small gaps (grey) and zeros in almost (up to numerical precision) every gap.}
    \label{fig:tme}
\end{figure}

Defining, $G_{N} = (\omega - M_{N})^{-1}$ (projected Green's function) and $V_N,W_N$ by the reduced singular value decomposition of the hopping matrix, $J_N = V_{N}D_{N}W_{N}^{\dagger}$, the TME reduces to (setting $t =1$) \cite{dwivedi2016bulk}
\begin{widetext}
\begin{eqnarray}\label{tmeq}
\overbrace{(W_{N}^{\dagger}G_{N}V_{N})^{-1}\begin{pmatrix}
1 & -(W_{N}^{\dagger}G_{N}W_{N})\\
V_{N}^{\dagger}G_{N}V_{N} & V_{N}^{\dagger}G_{N}W_{N}(W_{N}^{\dagger}G_{N}V_{N})- V_{N}^{\dagger}G_{N}V_{N}W_{N}^{\dagger}G_{N}W_{N}
\end{pmatrix}}^{\hat{T}_{q_{N},n}}
\begin{pmatrix}
V_{N}^{\dagger}\Psi_{n} \\ W_{N}^{\dagger}\Psi_{n-1}
\end{pmatrix} = 
\begin{pmatrix}
V_{N}^{\dagger}\Psi_{n+1} \\ W_{N}^{\dagger}\Psi_{n}
\end{pmatrix}.
\end{eqnarray}
\end{widetext}
When $W_{N}^{\dagger}G_{N}W_{N}\neq 0$ and $V_{N}^{\dagger}G_{N}V_{N}\neq 0$, $\hat{T}_{q_N,n}$ is unitary and has reciprocal eigenvalues, $\lambda_{T,1}\lambda_{T,2} = 1$. The spectrum, $E\in\Sigma$, is formed by energies for which $\left\vert\lambda_T\right\vert = 1$. By contrast, the spectral gaps, $E\in\mathbb{R}-\Sigma$, are the energies  for which $\left\vert\lambda_T\right\vert \in (0,1)\cup(1,\infty)$, see Fig.~\ref{fig:tme}. If $W_{N}^{\dagger}G_{N}W_{N}= 0$ or $V_{N}^{\dagger}G_{N}V_{N}=0$, $\hat{T}_{q_N,n}$ is no longer unitary and $\lambda_T = 0$ is possible.

In the limit $q_{N}\rightarrow\infty$, Rational approximate TMEs generate quasi-periodic eigenfunctions if the transfer matrix, $\hat{T}_{q_N,n}$, in Eq.~\eqref{tmeq} converges to the product of the full $2\times2$ ``unit-cell-free" quasi-periodic transfer matrices,
\begin{align}\label{tmeprodeq2}
\hat{T}'_{\alpha,n} = \prod_{i=1}^{q_N}\begin{pmatrix}
E - V\cos{(\Theta (n+i))} & -1 \\ 1 & 0 \end{pmatrix}.
\end{align}
Thus, we search for parameter values, i.e. $V<t$, where
\begin{align} \label{tmeprodeq}
\lim_{N\rightarrow\infty}\hat{T}'_{\alpha,n-q_{N}}\hat{T}'_{\alpha,n} = \lim_{N\rightarrow\infty} \hat{T}_{q_{N},n-q_{N}}\hat{T}_{q_{N},n}.
\end{align}
The full product of transfer matrices above corresponds to an infinite line of sites. Since, $\cos(\Theta x)$ has a recurrence point at infinity, the rational intra-unit cell terms in Eq.~\eqref{tmeq}, $G_{N}(\omega)$, are taken with periodic boundary conditions. Indeed, $G_{N}(\omega,\delta_y)$ is nothing but the projected Green's function, $G_{\perp,N}(\omega,k_{\parallel})$, previously used to establish topological bulk-boundary correspondence \cite{Borgnia2020,Slager2015,Rhim2018,Wilsons, mong2011edge}.

\sect{Projected Green's Function} 
The translation invariant intra-cell projected Green's function (pGF), $G_{N}(\omega)$ in Eq.~\eqref{tmeq}, is generated from a translation invariant Hamiltonian by inverse Fourier transforming the bulk Green's function back into real space along a single momenta, $G(\omega, x_{i},x_{f})$, and taking the component beginning and ending on the n-th unit cell, $x_{i} = x_{f} = n$.

For our 2D translation invariant rational approximates, we define the pGF as the zeroth inverse-Fourier mode with respect to a single momentum $k_{\perp}$ onto a single unit cell, 
\begin{align}\label{pgfeq}
\hat{G}_{\perp}(\omega,k_{\parallel},n)=\int dk_{\perp} G(\omega,k_{\parallel},k_{\perp})e^{ik_{\perp}\cdot n}.
\end{align}
In general, $k_{\parallel}, k_{\perp}$ represent momenta in parallel and transverse directions to the projection, and $n = 0$ can be chosen by translation invariance. Here, $G_N(\omega, \delta_y) = \hat{G}_{\perp,N}(\omega, k_{\parallel})$ with $\delta_y$ defined in Eq.~\eqref{2Dhamiltoniangauge}. 
Thus, convergence of rational TMEs reduces to the convergence of the rational Green's function, integrand of Eq.~\eqref{pgfeq}, to the irrational Green's function.

We note that zeros of the pGF arise from topological winding of the bulk Green's function in Eq.~\ref{pgfeq} \cite{Rhim2018,Slager2015,mong2011edge,Borgnia2020} and correspond to impurity bound states as for impurities of infinite strength -- edge modes, see S.I.~\ref{zerospolessi} and Ref. \cite{Slager2015,Borgnia2020,Rhim2018}. As illustrated in Fig.~\ref{fig:tme}, this generates a correspondence between pGF zeros and rank deficient energies for the transfer matrix-- red crosses indicate pGF zeros with the corresponding zero transfer matrix eigenvalue.  Thus, when rational pGFs converge to the irrational limit as discussed below, the band topology of the rational approximates can generate rank deficient points of the resulting irrational transfer matrix. These rank deficient points form topological obstructions for the spectrum (green dots in Fig.~\ref{fig:tme}). This highlights the need for 2D rational approximates which respect the quasi-periodic topology to insure the sequence of rational TMEs converges to the correct irrational TME.

\sect{Almost Mathieu Convergence} In addition to respecting the underlying quasi-periodic topology, the 2D rational approximates have a gauge choice absent from the 1D approximates. The magnetic unit cell can be chosen arbitrarily, but only certain choices will converge in the irrational limit, see Fig.~\ref{fig1} (center). 

Taking the AAH model as an example, for $V < t$ and almost every $\Theta$ (not Liouville), choosing a horizontal unit cell results in a convergent rational approximate pGF, while a vertical unit cell does not. For the AAH model we analytically prove convergence of certain unit cells given different choices of $V,t,\Theta$, S.I.~\ref{siboundstransfer}, but in the case of generic quasi-periodic Hamiltonians one can still numerically check the pGF convergence to find the appropriate unit cell in practice. 

We leave the mechanism behind the localization transition for a companion paper \cite{paper2}. For the purposes of this work, the existence of a convergent horizontal magnetic unit cell guarantees the 2D rational approximate TMEs converge to the irrational TME, and the horizontal unit cell implies the existence of extended solutions under projection back to 1D -- choosing a particular phase $\delta_y$, see Fig.~\ref{fig1}. These extended solutions are by construction invariant under phase shifts of the potential, ($\cos(\Theta x+\delta)\rightarrow\cos(\Theta x + \delta+\Theta)$, and are thus effectively translation invariant. In fact, Andre and Aubry originally called the metal-insulator transition a gauge symmetry breaking transition \cite{aubry1980annals,aubry1981bifurcation}, as metallic eigenfunctions are invariant under the $U(1)$ phase $\delta_y$, but this breaks down for the localized states. In this regard, the MIT is also reminiscent of an integrability breaking condition as the extended eigenfunctions have an extensive number of conserved quantities ($\delta_y$) which are broken under localization. Although, the mechanism is quite different as discussed in \cite{paper2}. 

Here we focus on finding the metallic parameter regime, and its breakdown. This is accomplished by  bounding (in a operator norm sense) the difference between the rational approximate pGFs and the irrational pGF, $\vert\vert G_{\perp,\alpha}(\omega)-G_{\perp,N}(\omega)\vert\vert$, generating a convergent sequence of rational approximate TMEs for the irrational TME. 

Bounding the pGF difference requires bounding the difference on a single unit cell between the bulk irrational Green's function, $G_{\alpha}(\omega,k_{\perp},k_{\parallel})$, and the bulk rational approximate Green's function, $G_{N}(\omega,k_{\perp},k_{\parallel})$ difference on a single unit cell, but Green's function poles make this non-trivial, S.I.~\ref{siboundstransfer}. One has to move away from the bulk spectrum (poles). Unfortunately, the rational Hofstadter Hamiltonian has exponentially small spectral gaps, as a function of $q_N$, making the Green's function eigenvalues exponentially large.

Instead, we shift off the real axis, $\omega\rightarrow\omega\pm i\epsilon$ and prove (S.I.~\ref{siboundstransfer}) that for any $\epsilon>0$, we can take $q_N$ sufficiently large such that for any $\delta>0$
\begin{eqnarray}\label{pgfdeltabound}
\vert\vert \re(G_{\perp,\alpha}(\omega\pm i\epsilon)-G_{\perp,N}(\omega\pm i\epsilon))\vert\vert &<& \delta\nonumber\\
\vert\vert \im(G_{\perp,\alpha}(\omega\pm i\epsilon)-G_{\perp,N}(\omega\pm i\epsilon))\vert\vert &<& \delta.
\end{eqnarray}
Or, 
\begin{eqnarray}\label{pgfepslim}
\lim_{\epsilon\rightarrow 0}\lim_{q_N\rightarrow\infty}\vert\vert G_{\perp,\alpha}(\omega\pm i\epsilon)-G_{\perp,N}(\omega\pm i\epsilon)\vert\vert = 0.
\end{eqnarray} 
As the Green's function is continuous on the upper and lower half of the complex plane, with the same limit, 
\begin{eqnarray}\label{pgfqNlim}
\lim_{q_N\rightarrow\infty}\vert\vert G_{\perp,\alpha}(\omega)-G_{\perp,N}(\omega)\vert\vert =0.
\end{eqnarray}
We discuss the analytic bound on the AAH model in S.I.\ref{siboundstransfer}, and shift focus to numerical applications.

\begin{figure}
    \centering
    \includegraphics[scale = .38]{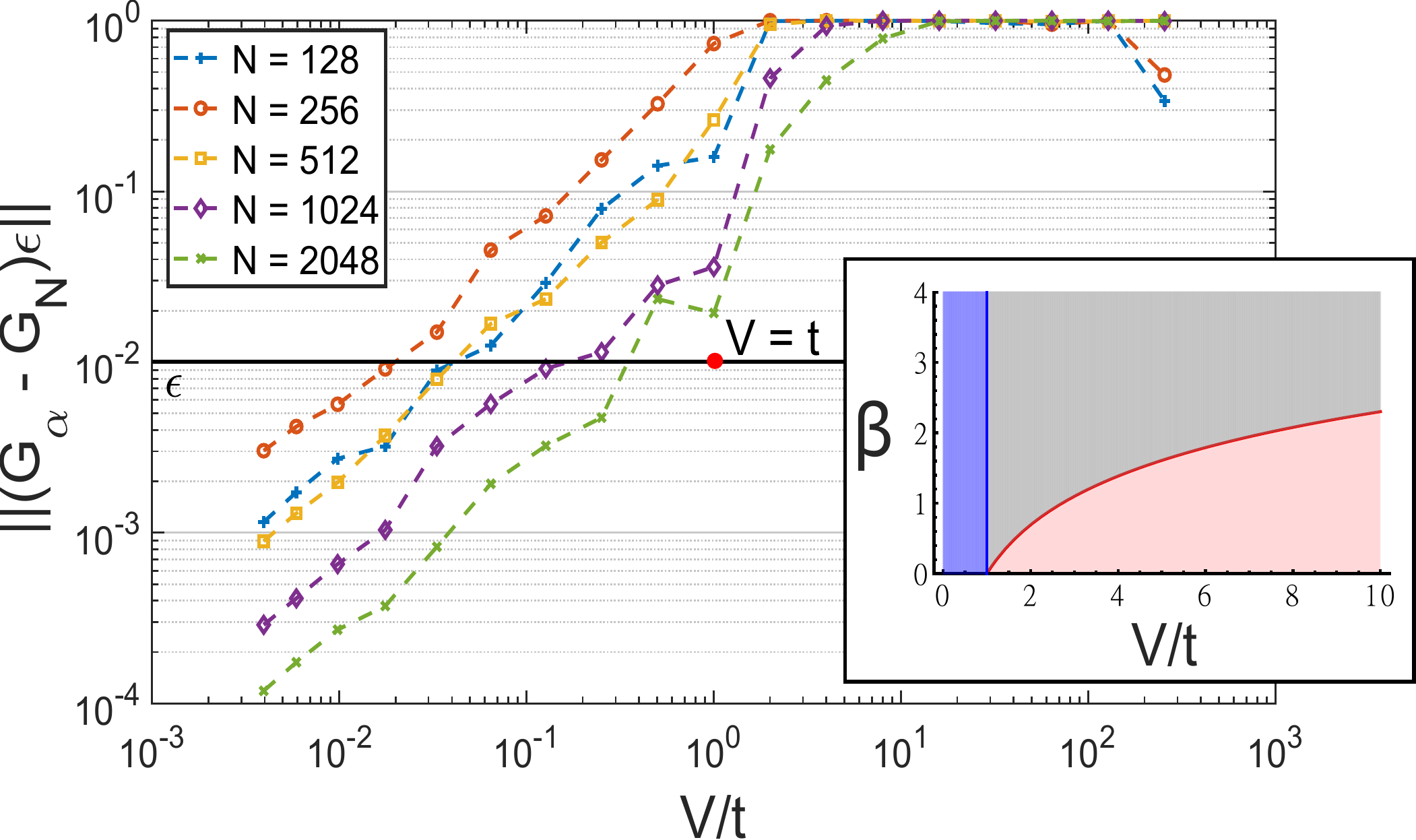}
    \caption{Plot of max (operator norm) difference over all $\omega$ between the rational and irrational pGFs for $\alpha = (\sqrt{5}-1)/2$. Convergence is defined by the ability to take $\vert\vert G_{\alpha}(\omega+i\epsilon)-G_N(\omega+i\epsilon)\vert\vert <\epsilon^{2}.$ For $V/t<1$ ($V/t>1$) horizontal (vertical) unit cell converges. Taking larger system sizes N, sharpens convergence/divergence [shrinking divergence for large $V/t$ caused by finite system size]. (inset) AAH phase diagram for Liouville $\alpha$, with $\beta$ the Liouville exponent and transition at $\beta = \ln{(V)}$ (blue: metal, red: insulator, gray: transition).}
    \label{fig2}
\end{figure}

\begin{figure*}
    \centering
    \includegraphics[scale=.28]{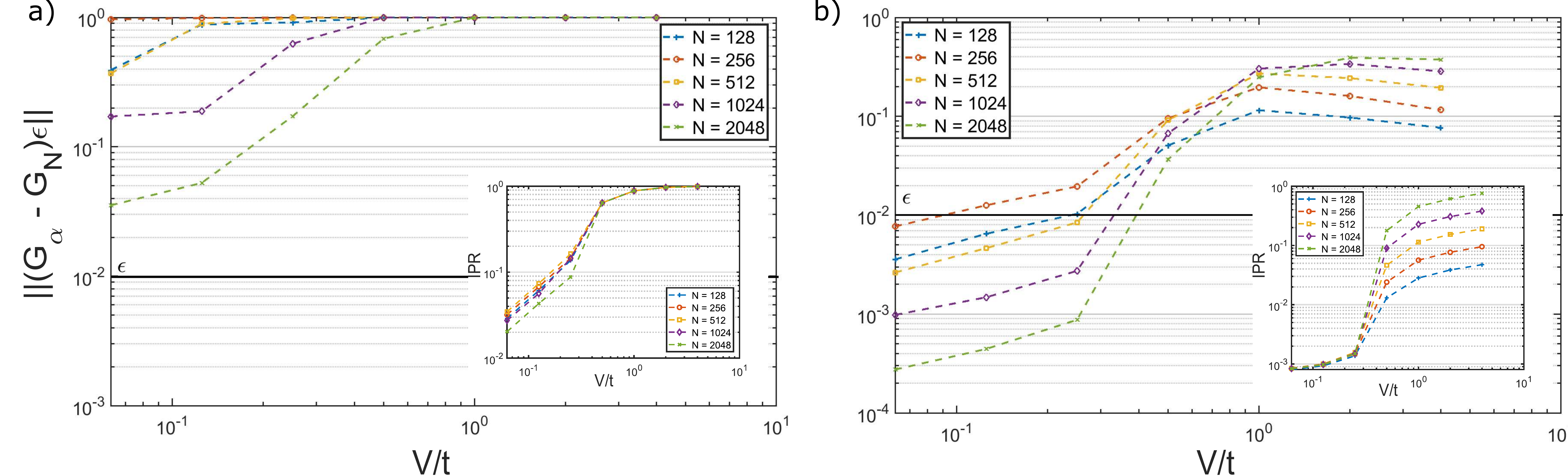}
    \caption{Convergence criteria versus the inverse Participation ratio (IPR) as measure of eigenfunction localization vs delocalization. Panel (a) shows max criteria of convergence vs inset max IPR. Panel (b) displays average criteria vs inset mean IPR. For non-self dual models, mobility edges persist deep into the metallic regime. Note, for small N, mean IPR doesn't converge to 1 as quasiperiodicity breaks down.}
    \label{fig:multimodal}
\end{figure*}

\sect{Self-Dual Examples}
We see a numerical validation of the analytic results for the AAH model in Fig.~\ref{fig2}. Note that for $V>t$ and a horizontal unit cell, as expected, the inequalities in Eq.~\eqref{pgfdeltabound} do not hold and the pGF difference does not converge, S.I.~\ref{siboundstransfer}. Similarly, choosing a vertical unit cell for $V>1$ ($V<t$ leads to the opposite bound (divergence), after a gauge transformation, S.I~\ref{siboundstransfer}. The localization mechanism for a vertical unit cell is deeply tied to quasi-periodic topology as we will report on in the near future\cite{paper2}.

Our convergence criteria is further validated for Louiville $\alpha$, where we recover the known result, often missed by numerical methods, that the MIT is shifted from $V =t$ to $V = te^{\beta}$ \cite{avila2017sharp,jitomirskaya1999metal,avila2006solving}. Here, $\beta$ is defined for any Liouville irrational number, $\alpha$, as the minimum number, $\beta(\alpha)$ for which $\vert\alpha -\frac{p_n}{q_n}\vert < e^{-\beta(\alpha)q_n}$. Thus, taking $\alpha$ Liouville and correspondingly $\delta = e^{-\beta q_{N}}$, our analytic bound in S.I.~\ref{siboundstransfer} becomes, 
\begin{eqnarray}
\vert\vert G_{\perp,\alpha}(\omega)-G_{\perp,N}(\omega)\vert\vert < (4V/\pi)^{q_N}e^{-\beta q_{N}}q_{N},
\end{eqnarray}
and the rational approximate pGFs converge for $V\lesssim e^{\beta}$. Thus, for $1<V<e^{\beta}$, both the horizontal and vertical unit cells are convergent and the localized phase is pushed to larger V \cite{avila2017sharp}. Note, while vertical unit cells also converge for $e^{-\beta}<V<1$, when $V<1$ the projection back to 1D is always well-defined by the duality transformation \cite{paper2}.

The AAH model is special for its sharp phase transition \cite{avila2017sharp}, but the convergence criteria above also captures \textit{mobility edges} -- states undergoing a localized to delocalized transition at a fixed energy -- that are absent in the AAH model. We demonstrate this with the generalized AAH model, parameterized by the onsite potential,
\begin{eqnarray}\label{potentialsi}
V(x) = 2V \frac{\cos(\Theta x+\delta_{x})}{1-b\cos(\Theta x+\delta_{x})},
\end{eqnarray}
where $b\in(-1,1)$ detunes the model from the AAH model. This model hosts a mobiity edge originating from the modified duality \cite{ganeshan2015nearest}, which depends on the energy $E$ of the eigenfunction in question
\begin{eqnarray}\label{GAAHduality}
b E = \textnormal{2 sgn}(V)(\vert t\vert-\vert V\vert).
\end{eqnarray}

We define the GAAH 2D parent Hamiltonian,
\begin{eqnarray}
    \mathcal{H}_{2D} &=& \sum_{x,\delta_x} \left[ t\hat{c}_{x+1,\delta_x}^{\dagger}\hat{c}_{x,\delta_x} +  h.c.\right]\nonumber\\
    &+&\left[g \sum_{r\geq0}e^{-\beta\vert r\vert} \cos{(\Theta x r +\delta_x)}\hat{c}^{\dagger}_{x,\delta_x}\hat{c}_{x,\delta_x} \right]
\end{eqnarray}
Notice the long-range hopping along the ``phase" coordinate and short range hopping along the ``real space" coordinate. As such, no simple gauge transformation will generate the 1D duality transformation in 2D. And, the rational approximates from S.I.~\ref{siboundstransfer} will only host finite unit cells for a horizontal unit cell choice. We can actually still analytically solve conditions for the horizontal unit cell convergence, S.I.~\ref{gaahsi}, finding for $\alpha$ diophantine, the pGF converges when $bE<2(t-V)$ with positive $t,V$. We can also see this with numerical methods Fig.~\ref{fig:gaah}.

Unlike in the AAH model, there is no well defined rational approximate to the vertical unit cell as hoppings are infinite range. As such, the topological arguments in \cite{paper2} do not hold and we expect a less sharp transition, seen numerically by \cite{ganeshan2015nearest,biddle2011localization}.

\sect{Breaking Self-Duality} It was noted early on in the study of quasi-periodic systems that duality was not driving the localization transition \cite{frohlich1990localization,aubry1980annals,dinaburg1975one}, and this can be seen clearly from the above formalism by choosing multi-modal quasi-periodic examples, i.e. taking the potential
\begin{eqnarray}
    V(x) = 2V(\cos{(\Theta x+\delta)}+\cos{(3\Theta x+\delta)}).
\end{eqnarray}

In Fig.~\ref{fig:multimodal} we see the failed horizontal unit cell convergence exactly aligns with the localization of eigenfunctions, even finding the mobility edge. While above examples are Nearest Neighbor (N.N.) models, the above tool can be generalized beyond N.N. by choosing a larger cut for the pGF \cite{Slager2015,Rhim2018,dwivedi2016bulk,mong2011edge}, a topic for future work.

\sect{Conclusions} Here we report on a generic tool for explaining the understanding and predicting the extended behavior of quasi-periodic eigenfunctions. Our framework is both boundary free and directly implementable for a wide class of models beyond the well studied AAH model. More importantly the technique is transparent, explaining a mechanism for the existence of metallic phases in quasi-periodic systems, even allowing for some analytic bounds. This transparency allows us to capture the shifted phase diagram of the AAH model for Liouville irrationals and establish clear regimes of validity, an important aspect of any numerical technique.

\begin{acknowledgments}
We thank Vir B. Bulchandani, Ruben Veressen, Matthew Gilbert, Nick G. Jones, Joaquin Rodriguez-Nieva, Dominic Else, Ioannis Petrides, Daniel E. Parker, Eitan Borgnia, Madeline McCann, Saul K. Wilson, Will Vega-Brown, and especially Matthew Brennan for insightful discussions.
AV was supported by a Simons Investigator award and by the Simons Collaboration on Ultra-Quantum Matter, which isa grant from the Simons Foundation (651440, A.V.).
R.-J.S. acknowledges funding from the Winton Programme for the Physics of Sustainability and from the Marie Sk{\l}odowska-Curie programme under EC Grant No. 842901 as well as from Trinity College at the University of Cambridge.
\end{acknowledgments}
\bibliography{Refs}

\clearpage
\newpage
\pagebreak
\onecolumngrid
\begin{center}
\textbf{\large Supplemental Information - Rational Approximations of Quasi-Periodic Probelms via Projected Green's Functions}
\end{center}
\setcounter{equation}{0}
\setcounter{figure}{0}
\setcounter{table}{0}
\setcounter{page}{1}
\renewcommand{\theequation}{S\arabic{equation}}
\renewcommand{\thefigure}{S\arabic{figure}}

\section{S.I. AAH Background}\label{measureAAHsi}
 The main results of this work generalize to multiple classes of quasiperiodic models, but the Andre-Aubry model (almost-Mathieu operator) is the most well studied. We introduce it and some background on current methods in the study of single-particle quasi-periodic models. The Hamiltonian is very simple, but presents a rich playground for new techniques in analysis and single-particle physics:
\begin{eqnarray}\label{HamiltonianAAHeqRS2}
\hat{H} = \sum_{x} t(\hat{c}^{\dagger}_{x+1}\hat{c}_{x}+\hat{c}_{x+1}\hat{c}^{\dagger}_{x})+2V\cos(\Theta x+\delta_{x})\hat{c}^{\dagger}_{x}\hat{c}_{x}.\quad
\end{eqnarray}
Here $\Theta = 2\pi a$, and we'll take $a\in\mathbb{R}-\mathbb{Q}$. Note, the case where $a\in\mathbb{Q}$ is called the Mathieu operator and is simply a periodic 1D band model.

\indent To understand why the almost-Mathieu operator is interesting to mathematics, beyond the very physically interesting metal-insulator transition, we introduce the notion of a spectral measure. For any self-adjoint linear operator, $T$, one can decompose its measure on the target Hilbert space, $\mathcal{H}$ as an \textit{absolutely continuous}, \textit{singularly continuous}, and \textit{pure point-like} components. The spectral measure of $T$ is defined with respect to a vector $h\in\mathcal{H}$ and a positive linear functional $f: T\rightarrow \bra{h}f(T)\ket{h} = \int_{\sigma(T)}fd\mu_{h}$, where $\sigma(T)$ is the spectrum of the operator T and $\mu_{h}$ is the unique measure associated with $h$ and $T$. The portion of the Hilbert space, i.e. the subspace of vectors, for which $\mu_{h}$ is dominated by the Lebesgue measure on the same subspace -- for every measureable set $A$, if the Lebesgue measure $L(A)= 0$, $\mu_{h}(A) = 0$ -- is absolutely continuous. By contrast, the pure point like component is the discrete portion of the spectrum where points can have finite measure in terms of $\mu_{h}$, but points have zero Lebesgue measure. The singularly continuous part of the spectrum is defined as the singular part of the spectrum -- the subspace which can be formed by a disjoint union of sets $A$ and $B$ for which $\mu_{h}(A) = 0$ when $L(B) = 0$ --  which is not pure point like.

\indent The original metal-insulator transition was shown non-rigorously through the duality of the Andre-Aubry under a Fourier transform-like operation, $\hat{c}_{k} = \sum_{x}\exp(i\Theta k x) \hat{c}_{x}$,
\begin{eqnarray}\label{HamiltonianAAHeqKS2}
{\tilde{\hat{H}}} = \sum_{k} V(\hat{c}^{\dagger}_{k+1}\hat{c}_{k}+\hat{c}_{k+1}\hat{c}^{\dagger}_{k})+ 2t\cos(\Theta k+\delta_{k}) \hat{c}^{\dagger}_{k}\hat{c}_{k}\quad,
\end{eqnarray}
see S.I.~\ref{ogargumentsi}. The model has a self-dual point for $V = t$, fixing a transition from momentum-like to position-like eigenfunctions. A more complete formulation of the problem, however, was constructed and proven for almost-Mathieu operators. It was proven that the spectrum of the almost-Mathieu operator is (setting $t = 1$)
\begin{enumerate}
    \item Absolutely continuous for all $\Theta$ and $\delta_{x}$ when $V < 1$
    \item Singularly continuous for all $\Theta$ and $\delta_{x}$ when $V = 1$
    \item Pure point-like for almost all $\Theta$ and $\delta_{x}$ when $V > 1$
\end{enumerate}
A pure point-like spectrum guarantees Anderson localization as it corresponds to eigenfunctions having finite measure at the eigenvalues and zero measure elsewhere. Thus, the eigenfunction is exponentially decaying on the lattice or is not normalizable. Or more formally, the pure-point like spectrum forces eigenfunctions to be Semi-Uniformly-Localized-Eigenstates \cite{deift1983almost}. By contrast, an absolutely continuous spectrum guarantees delocalization if the spectrum has finite measure, which has been shown to be the case for the Almost-Mathieu operator. Much less is known about the Singularly continuous case, and it has been the topic of multiple famous problems proposed by Barry Simon \cite{simon1984fifteen,simon2000schrodinger,simon2020twelve}. One of the few results on the Singularly continuous spectrum is its existence deep in the pure-point like regime for Liouville $a = 2\pi/\Theta$ -- sequence of rational approximates $\lbrace p_n/q_n\rbrace$ exists such that $\vert a - p_{n}/q_{n}\vert < n^{-q_{n}}$ \cite{last2007exotic}. In fact, for Liouville numbers, the pure-point like transition occurs for $\lambda = e^{\beta}$ with $\beta = \lim_{n\rightarrow\infty}\ln(q_{n})/q_{n+1}$ \cite{avila2017sharp}.

In this language, the almost-Mathieu operator becomes a clear bridge between the well understood Mathieu operator (periodic operators) and random disorder. Understanding localization for the almost-Mathieu operator directly links to our understanding of chaos and localization in disordered systems. And yet, we still do not understand the full parameter space of a 1D nearest neighbor hopping lattice model with a cosine potential. The almost-everywhere part of this problem is important as it determines the physical stability of the model. Modern techniques in the field rely on cocycle theory  \cite{avila2006reducibility,avila2009ten,jitomirskaya1999metal,bourgain2002continuity}, and the absolutely continuous part of the spectrum is conjectured to be equivalent to the almost-reduciblity of the corresponding cocycle  \cite{avila2006reducibility}. The connection with cocycle theory further highlights the importance of this problem, as the reducibility classes of $SL(2,\mathbb{R})$ cocycles are known to describe the onset of quantum chaos and directly link to the Lyapunov exponent \cite{avila2006reducibility,avila2006solving,bourgain2002continuity}.

\section{S.I. The Andre Aubry's Argument}\label{ogargumentsi}
Early studies of quasi-periodic system dynamics focused on the construction of eigenfunctions from sequences of rational approximates, inductively \cite{dinaburg1975one,frohlich1990localization}. While the original work by Andre and Aubry \cite{aubry1980annals} relied on the continuity of the Thouless parameter and self-dual models to explain the transition. The RG like induction methods proved rigorously the existence of a localized phase. For the AAH model\cite{dinaburg1975one,frohlich1990localization} and similar quasi-periodic potentials\cite{frohlich1990localization}, these methods demonstrated the emergence of a \textit{pure point-like} spectrum for strong enough onsite potentials (relative to hopping terms). A pure point-like spectrum enforces localized eigenfunctions as eigenfunctions lack support across any continuous energy windows \cite{dinaburg1975one,frohlich1990localization,jitomirskaya1998anderson,jitomirskaya1999metal,jitomirskaya2019critical}. Below we introduce the simple AAH model and note key insights about the breakdown of eigenfunction ergodicity.

\indent The original paper by Andre and Aubry \cite{aubry1980annals} rests on two fundamental requirements for a quasi-periodic Hamiltonian, its self-duality and its fidelity to a sequence of rational approximates. It proposed a Hamiltonian, the AAH model, which satisfies a self-duality constraint under a real-space to dual-space (momentum space in the continuum limit) transformation, $\hat{c}_{k} = \sum_{x}\exp(i\Theta k x) \hat{c}_{x}$:
\begin{eqnarray}\label{aahrealspacetop}
	\hat{H} = \sum_{x} t(\hat{c}^{\dagger}_{x+1}\hat{c}_{x}+\hat{c}_{x+1}\hat{c}^{\dagger}_{x})+2V\cos(\Theta x+\delta_{x})\hat{c}^{\dagger}_{x}\hat{c}_{x}\quad\\
	{\tilde{H}} = \sum_{k} V(\hat{c}^{\dagger}_{k+1}\hat{c}_{k}+\hat{c}_{k+1}\hat{c}^{\dagger}_{k})+2t\cos(\Theta k+\delta_{k})\hat{c}^{\dagger}_{k}\hat{c}_{k}\quad \label{aahkspacetop}
\end{eqnarray}
\indent Here $\Theta$ is some irrational parameter relative to $\pi$ and clearly for $t = V$ the Hamiltonian is self-dual, indicating the existence of a transition. One can introduce a sequence of rational approximates, $\lbrace a_{n}/b_{n}\rbrace_{n\in\mathbb{N}}$ with $a_{n},b_{n}\in\mathbb{Z}$ and $\lim_{n\rightarrow\infty} a_{n}/b_{n} = \Theta$. The sequence of Hamiltonians with periodic potentials links the density of states on either side of the duality transformation because there are well-defined bands. One can then write down the corresponding Thouless exponent for each side of the transition:
\begin{eqnarray}
\gamma(E) = \int_{-\infty}^{\infty}\log\vert E-E'\vert dN(E')\label{thouless_exp_gen}
\end{eqnarray}
For rational $a_{n}/b_{n}$, with $t = 1$ and $V = \lambda$, the transformation from Eq.~\eqref{aahrealspacetop} to Eq.~\eqref{aahkspacetop} takes $\tilde{V}\rightarrow1/\lambda$ and $E\rightarrow \tilde{E}/\lambda$, which implies $N_{\lambda,k} (E) = \tilde{N}_{1/\lambda,k}(E/\lambda)$ \cite{aubry1980annals}. So,
\begin{eqnarray}\label{thoulessduality}
\gamma(E) &=& \int_{-\infty}^{\infty}\log\vert \frac{E-E'}{\lambda}\vert d\tilde{N}\left(\frac{E'}{\lambda}\right)  +\log\vert\lambda\vert\nonumber\\
\gamma(E) &=& \tilde{\gamma}(\frac{E}{\lambda})+\log\vert\lambda\vert.
\end{eqnarray}
Since quasiperiodic systems do not have bands, but rather protected band gaps (discussed below \cite{jitomirskaya1998anderson,jitomirskaya1999metal,avila2011holder,zhao2020holder,amor2009holder,jitomirskaya2019critical}), the Thouless exponent must be non-negative by construction in 1D \cite{aubry1980annals}. Thus, for $\lambda>1$, $\gamma(E)>0$ and states are exponentially localized. This all relies on the continuity of the Thouless exponent, only proven in 2002 \cite{bourgain2002continuity}. Here, the 1D rational approximates differ drastically from the irrational limit, but the density of state is well described by the approximation. In fact, these spectral properties are topologically protected by the quasiperiodic pattern's robustness \cite{prodan2015}, further expanded below.

\indent Via the above arguments, Andre and Aubry demonstrated the existence of non-zero Thouless parameter for $V >1$. And, by the duality of the model, $\gamma(E)$ must be zero for $V<1$. This transition is unusually sharp, exhibiting exponential localization on either side due to the relation between the Thouless parameters of the self-dual models. Further, the methodology is quite general in 1D and can be extended to other self-dual models, even if the duality is energy dependent, see S.I.\ref{gaahexamplesi}. The argument breaks down in higher dimensions as the Thouless exponent is no longer guaranteed to be non-negative \cite{aubry1980annals}. 

\indent Returning to the Hamiltonian in Eq.~\ref{aahrealspacetop}, note the phase $\delta_{x}$ sets the "origin" of the pattern. The eigenvalues cannot depend on the phase $\delta_{x}$ in the thermodynamic limit. However, for $\lambda >1$, if a state of energy $E$ is localized to site $x$ when $\delta_{x} = 0$, then the state localized to site $x-\delta/\Theta$ has energy $E$ for the shifted Hamiltonian with phase $\delta_{x} = \delta$. Thus, the eigenfunctions of each eigenvalue do depend directly on the phase. In \cite{aubry1980annals,aubry1981bifurcation}, this is described as a gauge-group symmetry breaking transition. \\
\begin{widetext}
\section{S.I. Transfer Matrix}\label{transfermatrixsi}
The explicit construction of a $2\times2$ transfer matrix starts with a reduced SVD of the hopping matrix $J_{N}$,
\begin{eqnarray}
J_{N} &=& V_{N} D_{N} W^{\dagger}_{N}\\
J_{N}^{\dagger} &=& W_{N} D_{N}^{\dagger} V^{\dagger}_{N}
\end{eqnarray}
with $V^{\dagger}V = W^{\dagger}W = \mathbb{1}$ and $W^{\dagger} V = 0$, and in the case of $J_{N}$, 
\begin{eqnarray}
D_{N} = D_{N}^{\dagger}=\begin{pmatrix}
t & 0 & \ldots &0\\
0 & 0 & \ldots & 0\\
\vdots& \vdots&\ddots & \vdots\\
0 & 0 & \ldots & 0
\end{pmatrix},
\end{eqnarray}
Since our hopping matrix is rank 1, we truncate $D_{N} = t$ and and correspondingly, the $q_{N}\times 1$ dimensional operators
\begin{eqnarray}
W_{N}= \begin{pmatrix}
0 \\ \vdots \\ 0 \\(-1)^{q_{N}}
\end{pmatrix},
V_{N}= \begin{pmatrix}
(-1)^{q_{N}} \\0\\  \vdots \\ 0
\end{pmatrix}
\end{eqnarray}
Rewriting our unit cell Green's function as 
\begin{eqnarray}
G_{N}(\omega) = (\omega - M_{N})^{-1}
\end{eqnarray}
The transfer matrix equation reduces to
\begin{eqnarray}
\Psi_{n} &=& G_{N}J_{N}\Psi_{n+1} + G_{N}J_{N}^{\dagger}\Psi_{n-1}\\
\Psi_{n} &=& G_{N}V_{N}D_{N}W^{\dagger}_{N}\Psi_{n+1} +  G_{N}W_{N}D_{N}V^{\dagger}_{N}\Psi_{n-1}
\end{eqnarray}
Projecting into the $V_{N},W_{N}$ subspaces of $\Psi_{n}$,
\begin{eqnarray}
V^{\dagger}_{N}\Psi_{n} &=& V^{\dagger}G_{N}V_{N}D_{N}W^{\dagger}_{N}\Psi_{n+1} +  V^{\dagger}G_{N}W_{N}D_{N}V^{\dagger}_{N}\Psi_{n-1}\\
W^{\dagger}_{N}\Psi_{n} &=& W^{\dagger}G_{N}V_{N}D_{N}W^{\dagger}_{N}\Psi_{n+1} +  W^{\dagger}G_{N}W_{N}D_{N}V^{\dagger}_{N}\Psi_{n-1}.
\end{eqnarray}
This reduces the Transfer matrix equation to \cite{dwivedi2016bulk}, setting $t =1$,
\begin{align}
\begin{pmatrix}
(W_{N}^{\dagger}G_{N}V_{N})^{-1} & -(W_{N}^{\dagger}G_{N}V_{N})^{-1}(W_{N}^{\dagger}G_{N}W_{N})\\
(V_{N}^{\dagger}G_{N}V_{N})(W_{N}^{\dagger}G_{N}V_{N})^{-1} & V_{N}^{\dagger}G_{N}W_{N}- V_{N}^{\dagger}G_{N}V_{N}(W_{N}^{\dagger}G_{N}V_{N})^{-1}W_{N}^{\dagger}G_{N}W_{N}
\end{pmatrix}\begin{pmatrix}
V_{N}^{\dagger}\Psi_{n} \\ W_{N}^{\dagger}\Psi_{n-1}
\end{pmatrix} = 
\begin{pmatrix}
V_{N}^{\dagger}\Psi_{n+1} \\ W_{N}^{\dagger}\Psi_{n}
\end{pmatrix}
\end{align}
Notice all of the elements in the $2\times2$ transfer matrix are effectively scalars and thus commute with each other and we can just factor out the common factor $(W_{N}^{\dagger}G_{N}V_{N})^{-1}$,
\begin{eqnarray}
(W_{N}^{\dagger}G_{N}V_{N})^{-1}\begin{pmatrix}
1 & -(W_{N}^{\dagger}G_{N}W_{N})\\
V_{N}^{\dagger}G_{N}V_{N} & V_{N}^{\dagger}G_{N}W_{N}(W_{N}^{\dagger}G_{N}V_{N})- V_{N}^{\dagger}G_{N}V_{N}W_{N}^{\dagger}G_{N}W_{N}
\end{pmatrix}
\begin{pmatrix}
V_{N}^{\dagger}\Psi_{n} \\ W_{N}^{\dagger}\Psi_{n-1}
\end{pmatrix} = 
\begin{pmatrix}
V_{N}^{\dagger}\Psi_{n+1} \\ W_{N}^{\dagger}\Psi_{n}
\end{pmatrix}
\end{eqnarray}
When $W_{N}^{\dagger}G_{N}W_{N}\neq 0$ and $V_{N}^{\dagger}G_{N}V_{N}\neq 0$, $\hat{T}_{q_N,n}$ is unitary and has reciprocal eigenvalues, $\lambda_{T,1}\lambda_{T,2} = 1$. The spectrum, $E\in\Sigma$, is formed by energies for which $\left\vert\lambda_T\right\vert = 1$. By contrast, the spectral gaps, $E\in\mathbb{R}-\Sigma$, are the energies  for which $\left\vert\lambda_T\right\vert \in (0,1)\cup(1,\infty)$, see Fig.~\ref{fig:tme}. 

If ever $(V_{N}^{\dagger}G_{N}V_{N})=0$ and $(V_{N}^{\dagger}G^{2}_{N}V_{N})=0$ or $(W_{N}^{\dagger}G_{N}W_{N})=0$ and $(W_{N}^{\dagger}G_{N}^{2}W_{N})=0$ the transfer matrix has rank 1 and the only non-vanishing solution is localized at the edge. The corresponding energy is not in the bulk spectrum, $\omega\notin\Sigma$. The above condition happens when $G_{N}(\omega) = (\omega-M_{N})^{-1} = 0$. These zeros correspond to protected obstructions of the spectrum and are important to quasi-periodic localization \cite{paper2}.

\section{S.I. Bulk to Projected Green's Function}\label{siboundstransfer}
Our rational approximation to the irrational Hamiltonian follows the continued fraction approximation of irrational parameter $\alpha$ described above. 
We define,
\begin{eqnarray}
\hat{H}_{N} = \sum_{x}t(\hat{c}^{\dagger}_{x+1}\hat{c}_{x}+\hat{c}^{\dagger}_{x}\hat{c}_{x+1})+2V\cos{(2\pi \frac{p_{N}}{q_{N}}x+\phi)}\hat{c}_{x}^{\dagger}\hat{c}_{x}
\end{eqnarray}
for which we can define a $q_{N}$ site unit cell and Fourier transform into (setting $t =1$)
\begin{eqnarray}\label{fullmatrix}
\tilde{H}_{N} = \sum_{k}e^{ikx}c_{k}^{\dagger}c_{k}\begin{pmatrix}
2V\cos{(2\pi \frac{p_{N}}{q_{N}}1+\phi)} & 1 &0& \ldots & e^{ik}\\
1 & 2V\cos{(2\pi \frac{p_{N}}{q_{N}}2+\phi)} & 1& \ldots & 0\\
0 & 1 & 2V\cos{(2\pi \frac{p_{N}}{q_{N}}3+\phi)}& \ddots& 0\\
\vdots & \vdots &\ddots& \ddots  & \vdots\\
e^{-ik}  & \ldots  &0& 1 & V\cos{(2\pi \frac{p_{N}}{q_{N}}q_{N}+\phi)}
\end{pmatrix}.
\end{eqnarray}
The bulk Green's function is just the natural $G(\omega,k) = (\omega-\tilde{H}_{N}(k))^{-1}$ and the corresponding ``projected" Green's function is simply
\begin{eqnarray}
G_{\perp,N}(\omega)=\int \frac{dk}{2\pi}G(\omega,k)
\end{eqnarray}
We examine difference between the almost tri-diagonal matrix in E.Q.~\eqref{fullmatrix} and the irrational Hamiltonian on a given unit cell.
First, we set $\phi = 0$ w.l.o.g. and notice that 
\begin{eqnarray}
V\cos(2\pi\alpha x) = V\cos(2\pi(\frac{p_{N}}{q_{N}}+\delta_{N}) x) = V\cos(2\pi\frac{p_{N}}{q_{N}} x)- 2V\sin{(\pi\delta_{N}x)} \sin{(2\pi\frac{p_{N}}{q_{N}} x+\pi\delta_{N}x)}
\end{eqnarray}
So,
\begin{eqnarray}\label{diffbound}
\vert 2V\cos(2\pi\alpha x)-2V\cos(2\pi\frac{p_{N}}{q_{N}} x)\vert < \vert4V\sin{(\pi\delta_{N}x)}\vert = \vert 4V\sum_{n = 1}^{\infty} \frac{(-1)^{n}}{(2n-1)!} (\pi\delta_{N}x)^{2n-1}\vert<\vert 4V\pi\delta_{N}q_{N}\vert
\end{eqnarray}
where in the last inequality we used that an alternating and uniformly convergent series is bounded at any step and that $x\leq q_{N}$. Notice, that for any irrational number $\delta_{N}<\frac{1}{\sqrt{5}q_{N}^{2}}$ and thus
$$\vert 4V\pi\delta_{N}q_{N}\vert < \frac{4V\pi}{\sqrt{5}q_{N}}\rightarrow 0$$ in the limit of large $q_{N}$.
However, we keep $\delta_{N}$ going forward, to account for special cases of stronger bounds on $\delta_{N}$, i.e. Liouville numbers. Expanding the full irrational parameter on-site Green's function,
\begin{eqnarray}
G_{\perp,\alpha}(\omega)=\int\frac{dk}{2\pi}G_{N}(\omega)\left(\mathbb{1} + G_{N}(\omega)\left(\sum_{n=1}^{q_{N}}\left[2V\cos(2\pi\alpha n)-2V\cos(2\pi\frac{p_{N}}{q_{N}} n)\right]\ket{n}\bra{n}\right)\right)^{-1}
\end{eqnarray}
we see that
\begin{eqnarray}
G_{\perp,\alpha}(\omega)-G_{\perp,N}(\omega) &=& -\int\frac{dk}{2\pi}G_{N}^{2}(\omega)
\frac{\left(\sum_{n=1}^{q_{N}}\left[2V\cos(2\pi\alpha n)-2V\cos(2\pi\frac{p_{N}}{q_{N}} n)\right]\ket{n}\bra{n}\right)}{\left(\mathbb{1} + G_{N}(\omega)\left(\sum_{n=1}^{q_{N}}\left[2V\cos(2\pi\alpha n)-2V\cos(2\pi\frac{p_{N}}{q_{N}} n)\right]\ket{n}\bra{n}\right)\right)}.
\end{eqnarray}
We can bound the above expression in terms of the operator norm by using our above bound on
$$\vert 2V\cos(2\pi\alpha x)-2V\cos(2\pi\frac{p_{N}}{q_{N}} x)\vert<\vert4V\pi\delta_{N}q_{N}\vert.$$
Convergence then depends on the uniform convergence of 
\begin{eqnarray}
\left\Vert G_{N}(\omega)\left(\sum_{n=1}^{q_{N}}\left[2V\cos(2\pi\alpha n)-2V\cos(2\pi\frac{p_{N}}{q_{N}} n)\right]\ket{n}\bra{n}\right)\right\Vert^{k}<\Vert\left(G_{N}(\omega)4V\pi\delta_{N}q_{N}
\right)\Vert^{k}
\end{eqnarray}
as $k\rightarrow\infty$, where we have used the operator norm $\vert\vert A\vert\vert = \sup_{\psi\in \mathcal{H}} (\vert \vert A\psi\vert\vert/\vert\vert\psi\vert\vert)$. We need
\begin{eqnarray}
\Vert\left(G_{N}(\omega)4V\pi\delta_{N}q_{N}
\right)\Vert < 1
\end{eqnarray}
We have taken $\omega$ not in the spectrum, $\Sigma$ of the operator and $G_{N}(\omega)<\infty$, but this doesn't necessarily bound $G_{N}(\omega)<1/(\delta_N q_N)$. The operator norm of $G_{N}(\omega) = (\omega - E^{*})^{-1}$, or approximately the inverse of the gap width. The minimum gap width for the rational Hofstadter approximates is known, $\sim V^{q_N/2}$ \cite{jitomirskaya2020spectrum}, which means $G_{N}(\omega) \sim V^{-q_N/2}$. Taking this approach, there will always be gaps for which the pGF doesn't converge.

Instead, consider a small shift of $\omega$ towards the upper and lower half of complex plane $\omega\rightarrow\omega \pm i\epsilon$. This fixes the maximum of $G_{N}(\omega\pm i\epsilon) < \epsilon^{-1}$. We can always take a large enough $q_N$ such that $\delta_{N}q_{N} < \epsilon$ for any $\epsilon>0$. However, we now need both the imaginary and real parts of the offset pGF to converge.
\begin{align}
    \re{(G_{\perp,\alpha}(\omega\pm i\epsilon) - G_{\perp,N}(\omega\pm i\epsilon))} = -\int\frac{dk}{2\pi}\frac{\left[(\omega-\tilde{H}_{N})^{2} + (\omega-\tilde{H}_{N})(\tilde{H}_{N}-\tilde{H}_{\alpha})-\epsilon^{2}\right](\tilde{H}_{\alpha}-\tilde{H}_{N})}{((\omega-\tilde{H}_{N})^{2}+\epsilon^{2})^{2}\left(\mathbb{1}+\frac{2\omega (H_{N} - H_{\alpha})+\tilde{H}_{N}^{2}- \tilde{H}_{\alpha}^{2}}{((\omega-\tilde{H}_{N})^{2}+\epsilon^{2})}\right)}\\
    \im{(G_{\perp,\alpha}(\omega\pm i\epsilon) - G_{\perp,N}(\omega\pm i\epsilon))} = \int\frac{dk}{2\pi}\frac{\left[\pm 2i\epsilon(\omega-\tilde{H}_{N})+i\epsilon(H_{N} -\tilde{H}_{\alpha})\right](\tilde{H}_{\alpha}-\tilde{H}_{N})}{((\omega-\tilde{H}_{N})^{2}+\epsilon^{2})^{2}\left(\mathbb{1}+\frac{2\omega (H_{N} - H_{\alpha})+\tilde{H}_{N}^{2}- \tilde{H}_{\alpha}^{2}}{((\omega-\tilde{H}_{N})^{2}+\epsilon^{2})}\right)}
\end{align}
Now we use that $\vert\vert(\omega-\tilde{H}_{N})^{2}+\epsilon^{2}\vert\vert >\epsilon^{2}$, to cancel the bottom factors $\sim \frac{1}{\epsilon^{2}}$. And, we can use the bound from above, $\vert\vert\tilde{H}_{N} - \tilde{H}_{ch,\alpha}\vert\vert<2\pi\sqrt{\frac{1}{15q_{N}}}$ to get 
\begin{align}
   \vert\vert \re{(G_{\perp,\alpha}(\omega\pm i\epsilon) - G_{\perp,N}(\omega\pm i\epsilon))}\vert\vert < \int\frac{dk}{2\pi}\vert\vert\epsilon^{-4}\left[-\epsilon^{2}2\pi\sqrt{\frac{1}{15q_{N}}}+ (\omega-\tilde{H}_{N})\left(2\pi\sqrt{\frac{1}{15q_{N}}}\right)^{2} \right]\vert\vert\\
    \vert\vert\im{(G_{\perp,\alpha}(\omega\pm i\epsilon) - G_{\perp,N}(\omega\pm i\epsilon))}\vert \vert< \int\frac{dk}{2\pi}\epsilon^{-3}\left[\epsilon8\pi\sqrt{\frac{1}{15q_{N}}}+\left(2\pi\sqrt{\frac{1}{15q_{N}}}\right)^{2}\right].
\end{align}
Clearly the imaginary part will converge to zero for large $q_N$. We bound the real part by showing
\begin{eqnarray}\label{ourboundeq}
\vert\vert(\omega -\tilde{H}_{ch,N})\vert\vert < C\sqrt{q_N},
\end{eqnarray}
so that we can always take $q_N$ big enough to make $\left(2\pi\sqrt{\frac{1}{15q_{N}}}\right)^{2} C\sqrt{q_N}<\epsilon^{4}$ for any $\epsilon>0$ to get the uniform convergence above, for all Diophantine $\alpha$.

The operator norm obeys the triangle inequality, so
\begin{eqnarray}\label{triangleq}
\left\vert\vert\omega\vert- \vert\vert\tilde{H}_{N}\vert\vert\right\vert\leq\vert\vert(\omega-\tilde{H}_{N})\vert\vert \leq\vert\omega\vert+\vert\vert\tilde{H}_{N}\vert\vert
\end{eqnarray}
We can then  use the lower bound to prove divergence for $V>1$ and the upper bound to prove convergence for $V<1$. 

By \cite{jitomirskaya2019critical}, the semi-infinite spectrum of $\tilde{H}_{N}$ may contain up to 2 eigenvalues inside each gap; and up to 1 in each of the infinite intervals above and below the bulk. We need to bound the eigenvalues above and below the gap (fortunately symmetric). We do this by using Lagrange's inequality to bound the characteristic polynomial roots for $\tilde{H}_{N}$
\begin{align}
    \vert\vert\tilde{H}_{N}\vert\vert < 1+\max_{0<i<q_N}\left\lbrace \left\vert\frac{a_i}{a_{q_N}}\right\vert\right\rbrace
\end{align}
with the $a_i$ the polynomial coefficients, the largest being $a_0 = \det{\tilde{H}_N}$. It thus suffices to prove that the largest coefficient, the determinant is finite for some parameter regime of $\tilde{H}_N$. We do this via an arithmetic mean - geometric type mean argument (AM-GM). Consider without loss of generality, $t=1$, then 
\begin{eqnarray}
\det{\vert \tilde{H}_{N}\vert} &\leq& (\tr{\left\vert \frac{\tilde{H}_{N}}{q_{N}} \right\vert})^{q_N} =  \left\vert\sum_{x=1}^{q_N}\vert 2 V\cos{\Theta x}/q_{N}\vert\right\vert^{q_N}\nonumber\\
&\leq& \left(\frac{2V}{\pi} \int_{0}^{2\pi}\vert\cos{x} \vert dx\right)^{q_{N}} = (4V/\pi)^{q_N}.
\end{eqnarray}
Taking $V<\pi/4$, the bound holds and the rational transfer matrix approximates converge to the irrational transfer matrix. We extend this bound to $V < 1$ numerically ( intuitively seen by setting $\vert 2V \cos{x}\vert\rightarrow\max_x{\vert 2V \cos{x}\vert} = 2V$ to bound $\det{\tilde{H}_{N}}$). We leave the use of the lower bound in Eq.~\eqref{triangleq} to prove divergence for a companion paper \cite{paper2} and numerically show the divergence in Fig.~\ref{fig2}.

Thus, for $V<1$ and any $\epsilon>0$ both the imaginary and real parts of the rational pGFs converge for $q_N$ such that $\delta_{N}q_{N}<\epsilon^{4}$. This implies the uniform convergence of the rational approximates and the Green's function of the irrational limit is arbitrarily close to that of the translation invariant Green's function. Using the transfer matrix construction above, the quasi-periodic transfer matrix is arbitrarily close to the translation invariant transfer matrix. This is similar to almost-reducibility as defined by Artur Avilla in \cite{avila2006reducibility}.

Now, consider the case of $\delta_{N} = e^{-\beta q_{N}}$, when $\alpha$ is Liouville. Here,
\begin{eqnarray}
\Vert\left(G_{N}(\omega)4V\pi\delta_{N}q_{N}
\right)\Vert < \Vert\left(G_{N}(\omega)4V\pi e^{-\beta q_{N}}q_{N}
\right)\Vert 
\end{eqnarray}
Now, even if $V>1$ our bound holds,
\begin{eqnarray}
\det{\vert \tilde{H}_{N}\vert} \leq (4V/\pi)^{q_N}e^{-\beta q_{N}}q_{N},
\end{eqnarray}
and the rational approximate pGFs converge for $V\lesssim e^{\beta}$. Thus, for $1<V<e^{\beta}$, both the horizontal and vertical unit cells are convergent -- this reproduces results in \cite{avila2017sharp}. 
\end{widetext}
\subsection{S.I. Gauge Transformation}
If $V>1$, the above convergence proof fails and our rational Green's function approximates are no longer guaranteed converge to the irrational Green's function. Consequentially, the rational approximate transfer matrices no longer converge to the quasi-periodic transfer matrix and the quasi-periodic eigenfunctions are no longer the limit of the de-localized eigenufunctions of the rational approximate transfer matrices, defined by unit cells along the $x$-direction of the 2D parent Hamiltonian lattice. 

Instead we must apply a 2D gauge transformation to the 2D Hamiltonian, such that the ``magnetic" flux is contained in the $x$-direction and the unit cell is constructed in the $y$-direction. The resulting transfer matrix will be constructed from on-site projected Green's functions of the form
\begin{widetext}
\begin{eqnarray}
    G_{\perp,N} = \int\frac{d\phi}{2\pi}\left[\omega\mathbb{1}-\begin{pmatrix}
    2\cos{(2\pi \frac{p_{N}}{q_{N}}1+k_x)} & V &0& \ldots & Ve^{i\phi}\\
V & 2\cos{(2\pi \frac{p_{N}}{q_{N}}2+k_x)} & V& \ldots & 0\\
0 & V & 2\cos{(2\pi \frac{p_{N}}{q_{N}}3+k_x}& \ddots& \vdots\\
\vdots & \vdots &\ddots& \ddots  & V\\
Ve^{-i\phi}  & \ldots  &0& V & 2\cos{(2\pi \frac{p_{N}}{q_{N}}q_{N}+k_x)}    \end{pmatrix} \right]^{-1}.\nonumber
\end{eqnarray}
We can factor out a V from the entire equation above and generate the same Green's function as above with $V\rightarrow 1/V$ and $\omega\rightarrow \omega/V$. As a consequence, we have a clear convergence for $1/V<1$ or $V >1$ for the projected Green's function along a vertical unit cell in the limit of $q_N\rightarrow\infty$.

As a subtle point, while this gauge transformation is well defined in 2D for Liouville numbers, it fails to manifest in 1D, as the ``magnetic flux" can only shift to the hopping elements in $\hat{H}$ under a quasi-Fourier transformation. When the rational approximates in the horizontal unit cell converge, the quasi-Fourier transform is close (exponentially for $\alpha$ Liouville) to a rational Fourier transform, and we simply obtain the horizontal unit cell rational approximates. This point only manifests itself for Liouville choices of $\alpha$, such that the convergence regime extends into the convergence regime of the gauge transformed Green's function, $V>1$.
\end{widetext}
For Liouville $\alpha$, multiple gauge choices and unit cell configurations define convergent approximations to the full quasi-periodic projected Green's functions, but the in the projection to 1D is biased towards the horizontal unit cells as the eigenfunctions survive this projection and the ``gauge group symmetry" -- phase shift invariance -- is preserved \cite{paper2}.

\section{S.I. Projected Green's Function Formalism}\label{zerospolessi}
\indent In translation-invariant systems, the Brillouin zone allows for flexibility in writing locally computable formulas for topological invariants. In this language, Green's function zeros are singular and carry topological significance \cite{bernevig2013topological,Slager2015, Rhim2018, Borgnia2020, volovik2003universe, slager2019translational, Gurarie2011,mong2011edge}. More recently, it was noticed that bound state formation criteria along an edge are also defined by Green's function zeros \cite{Slager2015,Rhim2018,Borgnia2020,jitomirskaya2019critical,mong2011edge,volovik2003universe}, thereby tracking both topological invariants and their corresponding edge modes.

Extending this methodology beyond translation invariant systems consists of two steps. One must show both that Green's function zeros are still of topological significance and that edge formation criteria are still described by the presence of in-gap zeros. We first show the latter.

\indent The poles of the Green's function restricted to a particular site in position space correspond to an energy state at that particular site. Here restricted refers to the projection of the system Green's function, $G$, to a single site,
\begin{eqnarray}\label{gprojectsum}
G(\omega,\mathbf{r}_{\perp},\alpha_{\parallel}) = \sum_{\alpha} \vert\braket{\alpha\vert\mathbf{r}_{\perp}}\vert^{2}G(\omega,\alpha),
\end{eqnarray}
where $\alpha$ generically labels the Eigenvalues and $\alpha_\parallel$ is the remaining index post the contraction with $\mathbf{r}_{\perp}$. Generically, there will be many poles corresponding to the spectrum at $\mathbf{r}_{\perp}$, but they are not universal.
By adding on-site impurities and considering $G(\omega,\mathbf{r}_{\perp},\alpha_{\parallel})$ only in the band gap of the bare Green's function, any poles will be a result of the impurity potential, $\mathcal{V}(r) = \mathcal{V}\delta(r - \mathbf{r}_{\perp})$, binding a state in the gap. Then, by constructing an appropriate impurity geometry, and taking $|\mathcal{V}|\rightarrow\infty$, an edge is formed. Therefore, the condition for impurity localized states as $|\mathcal{V}|\rightarrow\infty$ is equivalent to the criteria for the formation of edge localized modes. And, impurity bound states correspond to zeros of the restricted in-gap Green's function. This is most readily seen by factoring the full Green's function, $G$ of some system with Hamiltonian $H_{0}$ and an impurity potential $\mathcal{V}$. That is, the full Green's function $G$ can be written in terms of the Green's function $G_{0} = (\omega - H_{0})^{\text{-}1}$ of the original system without the impurity, 
\begin{eqnarray}\label{factoring}
G(\omega,\alpha) &=& (\omega - (H_{0}+\mathcal{V}))^{\text{-}1}
=  (1+\mathcal{V}G_{0})^{\text{-}1}G_{0}.
\end{eqnarray} 
Correspondingly, impurity bound states (poles of $G$) in the gap (not a pole of $G_{0}$) must be a pole of $(1-\mathcal{V}G_{0})^{\text{-}1}$,
\begin{equation}\label{pgfdeteq}	\det \left[ G_{0}(\omega,\alpha)  \mathcal{V} - \mathbf{1} \right] = 0.
\end{equation}
For $|\mathcal{V}|\rightarrow\infty$, solutions require $G_{0} \rightarrow 0$, and the zeros of $G_{0}$ correspond to poles of $G$. Hence, the zeros of the restricted in-gap Green's function, $G(\omega,\mathbf{r}_{\perp},\alpha_{\parallel})$, correspond to edge modes, just as in the translation-invariant case \cite{Slager2015}.

\indent The above requires in-gap bound states of an aperiodic system to appear as zeros of the projected Green's function. We now derive the conditions under which such states are fixed by the pattern topology. In these cases, in-gap states survive small disorder \cite{prodan2015}, and impose constraints on system dynamics. The fundamental difference between the translation invariant and aperiodic cases comes down to the existence of a good momentum quantum number. For translation invariant systems, the $x$-basis is dual to the Brillouin zone momentum basis. Thus, each momentum eigenfunction is equally weighted in the projection, Eq.~\eqref{gprojectsum} with position $0$, and singular points of the projected Green's function directly relate to an obstruction of consistently writing the Green's function over $k$-space. Heuristically, for two bands separated by a topologically non-trivial gap, the eigenfunctions switch eigenvalues \cite{Slager2015,Borgnia2020,Rhim2018}. 

There is no such guarantee for generic aperiodic systems, but we can reduce the the constraints of translation invariance to a single condition for which the sum in Eq.~\eqref{gprojectsum} is in fact reduce-able to a sum over the topologically-fixed IDoS. We now focus on 1D systems (higher dimensions generalize by choosing codimension-1 surface \cite{Borgnia2020,Slager2015,Rhim2018}), where the projected Green's function on the site $x_{0}$ reduces to 
\begin{eqnarray}\label{1Dgprojectsum}
G_{\perp}(\omega,x_{0}) = \bra{x_{0}}\left[\sum_{\alpha}G(\omega,\alpha)\ket{\alpha}\bra{\alpha}\right]\ket{x_{0}},
\end{eqnarray}
with $\alpha$ indexing the eigenfunctions of the system. In 1D, it is clear that any shift of the projection site $x_{0}$ can be absorbed as a phase shift in the pattern. We illustrate this with the AAH model, whose Hamiltonian reads
\begin{align}\label{aahrealspace}
	\hat{H} = \sum_{x} t(\hat{c}^{\dagger}_{x+1}\hat{c}_{x}+\hat{c}_{x+1}\hat{c}^{\dagger}_{x})+2V\cos(\Theta x+\delta_{x})\hat{c}^{\dagger}_{x}\hat{c}_{x}.
\end{align}
Here $\Theta$ is some irrational multiple of $2\pi$ generating the quasiperiodic pattern. The dynamics are generated by translations $\tau_{x}$, taking $V\cos(\Theta x+\delta_{x})\rightarrow V\cos(\Theta (x+1)+\delta_{x})$. And, we can shift $x_{0}\rightarrow x$ by taking $\delta_{x} = \Theta(x-x_{0})$. We can clearly index the Hamiltonian (and corresponding Green's function) by the real-space phase $\delta_{x}$. This is a general property of quasiperiodic patterns, and the spectrum is invariant under this shift \cite{bourne2018non}. This is guaranteed by choosing the hull of the pattern to define our unital algebra, as the system dynamics guarantee any initial point can be translated into any other point on the hull. We can therefore rewrite Eq.~\eqref{1Dgprojectsum} as
\begin{eqnarray}\label{1Dgprojectsumavg}
G_{\perp}(\omega,x_{0}) = \frac{1}{N}\sum_{x}\bra{x}\left[\sum_{\alpha}G(\omega,\alpha)\ket{\alpha(\delta_{x})}\bra{\alpha(\delta_{x})}\right]\ket{x},\nonumber\\
\end{eqnarray}
with $\delta_{x} = -\Theta(x-x_{0})$ depending on $x$, and the eigenfunctions depend on the choice of phase, i.e. an eigenfunction localized at $x_{0}$ becomes localized at site $x$, but with the same energy and corresponding Green's function component, $G(\omega,\alpha)$. If this were a translation-invariant setting, it would be clear that the choice of $x_{0}$ cannot matter, and, thus the sum in Eq.~\eqref{1Dgprojectsumavg} must reduce to 
\begin{align}\label{1Dgprojectsum_nophase}
G_{\perp}(\omega,x_{0}) = \frac{1}{N}\sum_{\alpha}G(\omega,\alpha).
\end{align}
If such as for the AAH model with $V<t$, discussed above, translation invariance holds, then the quantization of the IDoS fixes the sum in Eq.~\eqref{1Dgprojectsum_nophase} and all states are summed over. As a consequence, for an $\omega$ in a spectral gap, states above and below the gap contribute to the sum, such that for some $\omega_{*}$, $G(\omega)\vert_{\omega =\omega_{*}} = 0$. In translation invariant systems, the location of $\omega_{*}$ is usually protected by symmetries such as chirality, fixing states to be at equal energies above and below the gap. In quasiperiodic systems, the IDoS fixes the number of states above and below the gap\cite{bourne2018non}. For some gap labeled, $F$, the integrated density of states (IDoS) below each gap is fixed, i.e. for the AAH model $\text{IDoS}(F) = (m+n\Theta)\cup[0,N]$ \cite{bourne2018non}. Thus, for each gap, $F$, the relevant $\omega_{F}$ for which $G(\omega\in F)\vert_{\omega = \omega_{F}} = 0$ would also be fixed. This observation motivates our generalization to aperiodic systems.

\section{S.I. Topological Criterion} \label{topologicalchoiceSI}
As discussed in the main text, the transition is completely constrained by the projected Green's function zeros, forcing a sharp binary choice between the abs. continuous and pure point like spectra. 

We present a short review on the origins of non-commutative topological invariants in quasiperiodic systems. Not only do these systems exhibit a non-vanishing strong 1D topological invariant, it's existence consequentially dominates the bulk dynamics.
\begin{widetext}
\subsection{S.I. AAH Algebra}\label{siaahalgebra}
We follow the work of Prodan \cite{prodan2015} in deriving explicit topological invariants in the AAH model context. We construct the unital algebra and use it to label the resulting spectral gaps of the AAH Hamiltonian. Recall that it reads
\begin{eqnarray}\label{1Dham}
\mathcal{H}_{\delta_{x}} =\sum_{n}t \hat{c}_{n+1}^{\dagger}\hat{c}_{n} + \textnormal{h.c.} + 2V\cos(\Theta n+\delta_{x})\hat{c}_{n}^{\dagger}\hat{c}_{n},
\end{eqnarray}
in terms of the creation operators $c^{\dagger}_i$, lattice constant $a$, potential $V$ that depends on the position $n$ and is indexed by phase $\delta_{x}$.
The model exhibits a duality under the pseudo-Fourier transformation $c_{k} = \sum_{n} \exp(-ikn)c_{k}$  \cite{aubry1980annals}. Considering, $\delta_{x} = 0$, one obtains 
\begin{eqnarray}
    \tilde{\mathcal{H}}(k) &=& \sum_{k,k',n}t e^{ik(n+1)-ik'n}\hat{c}_{k}^{\dagger}\hat{c}_{k'}+t^{*} e^{ikn-ik'(n+1)}\hat{c}_{k}^{\dagger}\hat{c}_{k'} + V(e^{2\pi i an +i(k-k')n}+e^{-2\pi i an+i(k-k')n})\hat{c}_{k}^{\dagger}\hat{c}_{k'}\nonumber\\
  \tilde{\mathcal{H}}_{\phi}  &=& \sum_{k} 2t\cos(\Theta k)\hat{c}_{k}^{\dagger}\hat{c}_{k} + V(\hat{c}_{k+1}^{\dagger}\hat{c}_{k} +\textnormal{h.c.}).
\end{eqnarray}
where in the last line we have set $k = \Theta m$ and defined $\sum_{n} \exp(i\Theta n(m-m')) = \delta(m-m')$ in the limit $n\rightarrow\infty$.  
A natural equivalence emerges between $\mathcal{H}$ and $\tilde{\mathcal{H}}$ under $V\rightarrow t$, implying the model undergoes a transition for $V = t$, being the well known 1D metal-insulator transition. 
Considering the limits $V = 0$ and $t = 0$, the duality relates extended (momentum-localized) eigenfunctions to position-localized states. 

\indent The duality in the AAH model has been focus of many localization studies, past and present \cite{aubry1980annals,jitomirskaya2019critical}. The model took on new light, however, when \cite{kraus2012topological} noticed it could be parameterized by the phase choice $\delta_{x}$ \cite{kraus2012topological,jitomirskaya2019critical,avila2006solving,bellissard1982quasiperiodic}. Naively, this phase choice is irrelevant as it corresponds to a shift in initial position of an infinite chain, but the 2D {\it parent} Hamiltonian, as function of $x$ and  $\delta_{x}$, has a topological notion. In particular, it corresponds to a 2D tight-binding model with an irrational magnetic flux per plaquette. Explicitly,
\begin{eqnarray}\label{2Dhamiltonian2}
    \mathcal{H} &=&\sum_{n,\delta_{x}}t \hat{c}_{n+1,\delta_{x}}^{\dagger}\hat{c}_{n,\delta_{x}} + t^{*} \hat{c}_{n,\delta_{x}}^{\dagger}\hat{c}_{n+1,\delta_{x}} + 2V\cos(\Theta x+\delta_{x})\hat{c}_{n,\delta_{x}}^{\dagger}\hat{c}_{n,\delta_{x}},\nonumber\\
   \tilde{\mathcal{H}}  &=&\sum_{n,m,m'}t \delta_{m,m'}(\hat{c}_{n+1,m}^{\dagger}\hat{c}_{n,m'} + t^{*} \hat{c}_{n,m}^{\dagger}\hat{c}_{n+1,m'} )+V\left(e^{i\Theta x}\delta_{m+1,m'}+e^{-i\Theta x}\delta_{m-1,m'}\right)\hat{c}_{n,m}^{\dagger}\hat{c}_{n,m'},\nonumber\\
    \tilde{\mathcal{H}}  &=&\sum_{n,m}t (\hat{c}_{n+1,m}^{\dagger}\hat{c}_{n,m} + \hat{c}_{n,m}^{\dagger}\hat{c}_{n+1,m} )+ V(e^{i\Theta n}\hat{c}_{n,m+1}^{\dagger}\hat{c}_{n,m} + e^{-i\Theta n}\hat{c}_{n,m-1}^{\dagger}\hat{c}_{n,m}).
\end{eqnarray}
The 2D spectrum amounts to a Hofstadter butterfly when varying the flux per plaquette, $\Theta$. For any rational flux, $\Theta/2\pi = p/q \in\mathbb{Q}$, one can define a magnetic unit cell specifying bands that have a Chern number, which sum to zero. This is however not possible for an irrational flux. In this case strategies outlining sequences of rational approximates, with similar band gaps, to find topological invariants were employed \cite{PhysRevB.91.014108}. 

\end{widetext}
\indent Hamiltonian \eqref{2Dhamiltonian2} is manifestly topological for all rational fluxes, $a$ \cite{PhysRevB.91.014108}. We can therefore create a sequence of rational approximates to an irrational flux $\lbrace a_{n}\rbrace$, such that $\lim_{n\rightarrow\infty} a_{n} = a_{*}$. The problem, however, arises when projecting back down into one dimension. This is made most clear by considering the spectrum for different phase choices $\delta_{x}$, contrasting the irrational and rational case. For example, for $a = 1/2$, the choice $\delta_{x}$ changes the maximum amplitude of the on-site potential. In the irrational case, however, it has no effect on the spectrum and acts as a translation. Hence, the projection of the sequence of rational approximates {\it does not} create a 1D sequence of rational approximates to the AAH model \cite{prodan2015}. Instead more recent works leverage powerful tools from non-commutative geometry to tackle the problem conclusively, finding deep connections between non-commutative topological invariants and the inherited topology for the 1D projection, i.e the AAH model. In fact, methods described below describe both the inheritance of a 2D topological invariant and a bulk-boundary correspondence in the model \cite{prodan2015}. 

\begin{figure*}
    \centering
    \includegraphics[scale =.1]{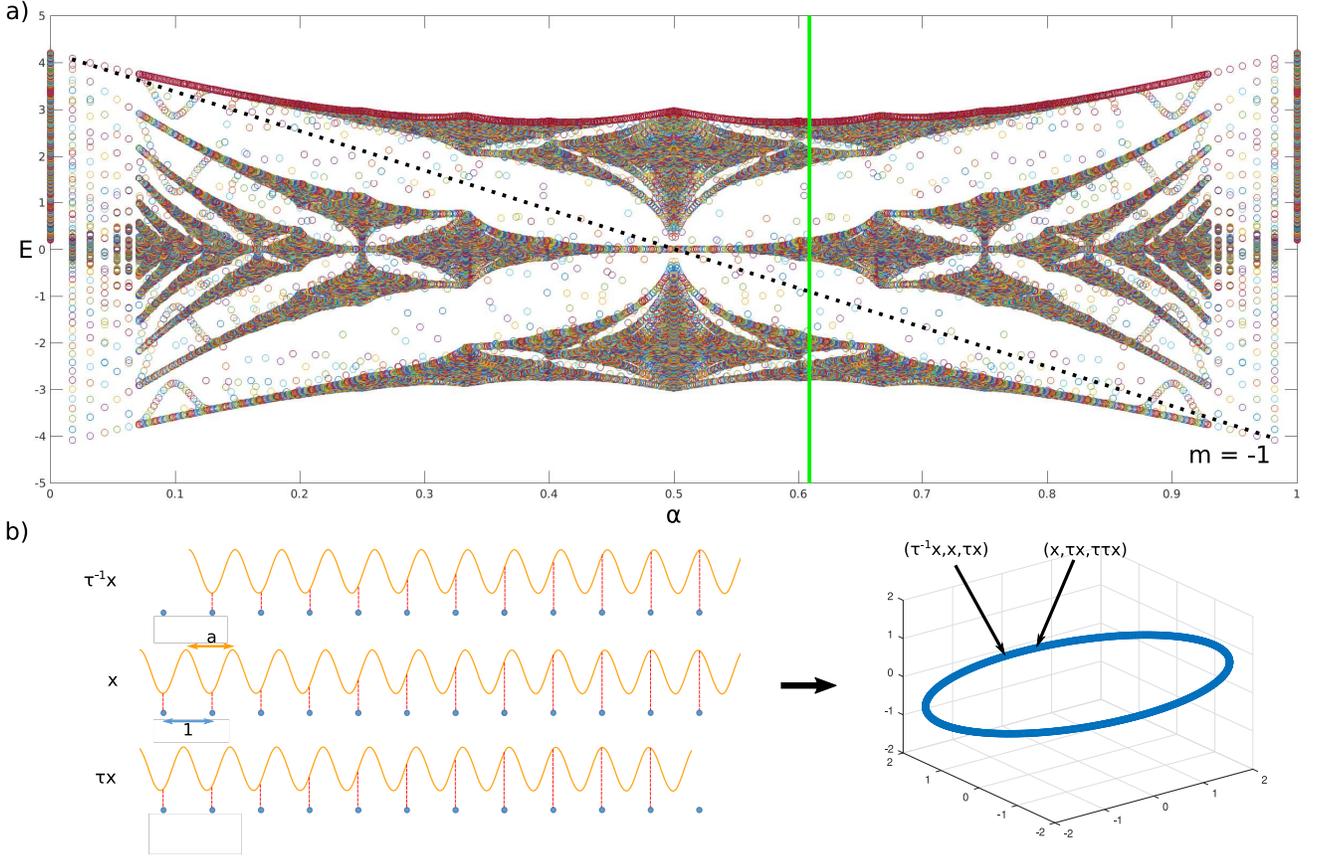}
    \caption{(a) quasiperiodic spectrum as function of plaquette flux $\Theta = 2\pi\alpha$. Also note the gap labeling index corresponds to slope of line $y = m\alpha + n$ with $m,n\in\mathbb{Z}$. (b) Illustration of quasi-periodic pattern generating a minimal surface (hull). This forms the underlying unital algebra, taking the place of a Brillouin zone.}
    \label{patternfigure}
\end{figure*}

\subsubsection{Non-commutative topological characterization}  We formalize the ideas behind a parent Hamiltonian for the AAH model and show the AAH model obeys the same algebra as the Hofstader Hamiltonian (defined below), forming a non-commutative torus. We show these two Hamiltonians are equivalent up to representation and all topological properties carry over \cite{prodan2015}.

\indent Consider the translation operator, acting on $\mathcal{H}_{\delta_{x}}$ as
\begin{align}\label{xspacetranslation}
    T^{n}\mathcal{H}_{\delta_{x}}T^{\dagger n} = \mathcal{H}_{\delta_{x}+n\Theta}.
\end{align}
For $\Theta/2\pi\notin\mathbb{Q}$, the repeated action of the translation operator parameterizes the motion of the  phase, $\phi\in\mathbf{S}$, along the unit circle. On the set of continuous functions over $\mathbf{S}$, $C(\mathbf{S}): \mathbf{S}\rightarrow\mathbb{C}$, the translation operator acts similarly. That is, for some $f \in C(\mathbf{S})$,
\begin{align}
    \alpha_{n}:f(\phi) \rightarrow f(\phi+\Theta n).
\end{align}
More formally, one paramterizes the action of $\mathbb{Z}$ on $\mathbf{S}$ as a dynamical system and then constructs its dual, i.e. the pattern hull described in the main text. The key step, however, is to define a unitary which acts in place of the translation operator on functions $f\in C(\mathbf{S})$.
\begin{align}
    u^{n}f(\phi)u^{-n} = f(\phi+\Theta n)
\end{align}
In this language, we can define elements of the space $C(\mathbf{S})\rtimes_{\alpha}\mathbb{Z}$, the semi-direct product between complex continuous functions of $\mathbf{S}$ and the translations $\mathbb{Z}$ generated by $\alpha: f(\phi) \xrightarrow{\alpha_{n}} f(\phi+\Theta n)$ as
\begin{align}
    \mathbf{a} = \sum_{n\in\mathbb{Z}} a_{n}u^{n}.
\end{align}
In the above $a_{n} \in C(\mathbf{S})$, and $u^{n}$ corresponds to a translation along $\mathbb{Z}$. The main benefit of defining this operator algebra corresponding to the 1D translations is that we can pick a representation of the Hilbert space,
\begin{align}\label{aahrep}
    \pi_{x}(\mathbf{a}) = \sum_{n,x\in\mathbb{Z}}a_{n}(\delta_{x}+\Theta x)\ket{x}\bra{x}T^{n},
\end{align}
with $\phi \in [0,2\pi]$. In this representation, the elements, $\mathcal{H}_{\delta_{x}}$, are simply represented as $\pi_{\phi}(\mathbf{h})$. Here, 
\begin{align}\label{Celement}
    \mathbf{h} = u+u^{-1} + 2V cos(\phi)
\end{align}
is an element of $C(\mathbf{S})\rtimes_{\alpha}\mathbb{Z}$, and the potential shifts when $T$ acts on $\phi$ at each site, i.e. $(\phi + \Theta n)\textnormal{mod}_{2\pi}$.

\indent As done for Eq. \eqref{2Dhamiltonian2}, we now show that this element $h\in C(\mathbf{S})\rtimes_{\alpha}\mathbb{Z}$ also generates the Hofstader Hamiltonian, see also \cite{prodan2015}. We first rewrite the Hofstader Hamiltonian as
\begin{align}\label{hofstaderhamiltonian2d}
    H_{\Theta} = \sum_{x,y} T_{x}+T_{x}^{-1}+V(T_{y}+T_{y}^{-1}),
\end{align}
where 
\begin{eqnarray}
T_{x}\ket{x,y} = \ket{x+1,y} \ \ \textnormal{and} \ \ 
T_{y} \ket{x,y} = e^{-i\Theta x}\ket{x,y+1}\nonumber
\end{eqnarray}
are magnetic translations with commutation relations $T_{x}T_{y} = e^{i\Theta}T_{y}T_{x}$. In this form, we define unitary operators $u$, as before, and $z = \exp{i\phi}$ acting on $C(\mathbf{S})$ corresponding to the translations along $x$ and $y$, respectively. These have the same commutation relations $uz = e^{i\Theta}zu$ and allow for a representation of the Hilbert space of $l^{2}$-normed functions on $\mathbb{Z}^{2}:$
\begin{align}\label{2delements}
    \pi' (\textbf{a}) = \sum_{n,m} f_{n}T_{x}^{n}T_{y}^{m}.
\end{align}
Then, $H_{\Theta} = \pi'(\textbf{h})$, for 
\begin{align}
    \mathbf{h} = u+u^{-1} + V(z+ z^{-1}) = u+u^{-1} + V (e^{i\phi} + e^{-i\phi})
\end{align}
as in Eq. \eqref{Celement}. 
\indent Framing the problem in terms of the operator algebra $C(\mathbf{S})\rtimes_{\alpha}\mathbb{Z}$ allows one to use techniques from non-commutative geometry to solve the problem immediately. In particular, this is a unital *-algebra for which a non-commutative calculus can be defined. The elements of the algebra are
\begin{align}
    \mathbf{a}  = \sum_{m,n\in\mathbb{Z}}f_{m,n}z^{m}u^{n},
\end{align}
\indent We can further revert to tools from non-commutative geometry to compute topological invariants. Details can be found in \cite{prodan2015}. We simply state the results here.
One can define differentiation intuitively along each direction:
\begin{eqnarray}
\partial_{1}\mathbf{a} &=& i\sum_{m,n\in\mathbb{Z}}m f_{m,n}z^{m}u^{n},\nonumber\\
\partial_{2}\mathbf{a} &=& i\sum_{m,n\in\mathbb{Z}}n f_{m,n}z^{m}u^{n}.
\end{eqnarray}
And, then integration follows as the inverse operation, $\mathcal{I}(\mathbf{a}) = f_{00}$, i.e. the constant term. These operations along with the algebra define the non-commutative Brillouin torus \cite{bellissard1986gaplabeling,bellissard1986k,bellissard2000hull}, $(C(\mathbf{S})\rtimes_{\alpha}\mathbb{Z},\partial,\mathcal{I})$, and form a special case of a spectral triple.

\indent Thus far, this section has only been a formalization of the concepts explained above. However, expressing the system as a spectral triple allows us to bring down the hammer of non-commutative geometry. In particular, there has been a careful formulation of K-theory in the case of spectral triples \cite{bellissard1986k,bourne2018non}. For the particularly simple case of a non-commutative torus, one can write down a locally computable index formula \cite{bourne2018non,bellissard1982quasiperiodic,bellissard1986gaplabeling,bellissard2000hull,bellissard1986k}, and compute a Chern number.
We introduce a projection operator, $\mathbf{p} = 1/2(1+\sgn(\epsilon_{F}-\mathbf{h}))$, which defines a filling of the spectrum below some Fermi level, $\epsilon_{F}$. The first non-commutative Chern number is then given by\cite{prodan2015,bellissard1986gaplabeling,bellissard1986k,bellissard2000hull,prodan2013non}
\begin{align}
    \textnormal{Ch}_{1}(\mathbf{p}) = 2\pi \mathcal{I}(\mathbf{p}[\partial_{1}\mathbf{p},\partial_{2}\mathbf{p}]).
\end{align}
Like the normal Chern invariant, this is well defined as long as there is a finite spectral gap.
Here we replicate the key result of \cite{prodan2015}, by computing this integral in the representation given by Eq. \eqref{aahrep}.
where $\mathcal{I}(\mathbf{a}) = 1/(2\pi)\int_{\mathbf{S}}d\phi f_{0}(\phi)$. In this simple case, the integral reduces to 
\begin{align}
    \mathcal{I}(\mathbf{a}) &=& \lim_{N\rightarrow\infty} \frac{1}{2N}\sum_{-N\leq x\leq N}f_{0}(\phi+\Theta x) =  \tr_{L}(\pi_{\phi}(\mathbf{a}))
\end{align}
where we have used that $f_{0}(\phi+\Theta x) = \bra{x}\pi_{\phi}(\mathbf{a})\ket{x}$, and $\tr_{L}$ is the normalized trace. In this representation
\begin{align}
    \textnormal{Ch}_{1} = 2\pi i \tr_{L}\left(\pi_{\phi}(\mathbf{p}[\partial_{1}\mathbf{p},\partial_{2}\mathbf{p})\right)
\end{align}
Using that, 
\begin{eqnarray}
\pi_{\phi} (\partial_{1}\mathbf{p})=\partial_{\phi}\pi_{\phi}(\mathbf{p}) \ \ \textnormal{and} \ \ \pi_{\phi} (\partial_{2}\mathbf{p})=i\left[X,\pi_{\phi}(\mathbf{p})\right],\nonumber
\end{eqnarray}

Defining $P_{\phi} = \pi_{\phi}(\mathbf{p})$, the projection operator in the AAH representation, the Chern number takes on a simple form,
\begin{align}
    \textnormal{Ch}_{1} = -2\pi  \tr_{L}\left(P_{\phi}[\partial_{\phi}P_{\phi},[X,P_{\phi}]]\right).
\end{align}
\indent Therefore, the Hofstader and AAH Hamiltonians are generated by the same element of $C(\mathbf{S})\rtimes_{\alpha}\mathbb{Z}$. This proves that the topological invariant of the 2D Hofstadter Hamiltonian is inherited by the 1D AAH model and explains the natural appearance of bulk-boundary correspondence -- the existence of boundary localized states reflecting the bulk topological invariant. The topological invariant is robust to disorder, and the edge spectrum is gapless when cycling through $\phi$ \cite{prodan2015}. 

An interesting consequence of this pGF transfer matrix formalism is the clear connection via the transfer matrix poles between the non-commutative geometry of the system and the spectral measure. Interestingly, this topological criteria was also noticed by \cite{jitomirskaya2019critical} using a different set of techniques to analyze the semi-infinite AAH model.

\subsection{S.I. Gap-Labeling Theorems}\label{gltheoremsi}
A well studied question arises from this topological criterion. Non-commutative geometry predicts the existence of gaps in the quasi-periodic integrated density of states (IDoS) depending on the irrational parameter $\alpha$ in the AAH Hamiltonian. Given the clear role topology plays in the dynamics, it was predicted and shown that these gaps in the IDoS form open sets in the complement of the quasi-periodic spectrum \cite{bellissard1986gaplabeling,bellissard1982quasiperiodic,bellissard1986k}.

Quasi-periodic systems can be indexed by patterns generating a "deterministic" disorder \cite{bourne2018non,bellissard2000hull,prodan2013non}. In Fig.~\ref{patternfigure}b peaks can be labeled by a coordinate, $P = \lbrace p_{i}\rbrace_{i\in\mathbb{Z}}$, forming a pattern. In the absence of a Brillouin zone, we consider the pattern as a dynamical system and find its convex hull - the minimal surface into which it can be embedded, $\Omega$. For example, for a simple generator such as $G = \cos(\Theta x)$ with $\Theta/2\pi \notin\mathbb{Q}$ it forms a ellipse. However, we can similarly find the hull for more complicated patterns upon defining the map, $f = \lbrace p_{i} \in P\vert f(p_{i}) = (p_{i+1} - p_{i},p_{i+2}-p_{i+1},\ldots)\in X\rbrace$ where $X$ is a hyper-cube of edge length defined by the pattern $P$. Although each element of the pattern is assigned a coordinate in an arbitrarily high dimensional space, there are only as many linearly independent coordinates as there are generators of the pattern. For the aforementioned sinusoidal generator of fixed amplitude, the element $p_{i+1} - p_{i}$ sets the period and further elements - $p_{i+2}-p_{i+1},\ldots$ - are linearly dependent on the first two by a translation. Consequently, it forms the anticipated ellipse in any dimensional hypercube rather than a higher dimensional surface, see Fig.~\ref{patternfigure}b. Ergodicity on this minimal embedding implies there exists a trajectory between any initial approaching (arbitrarily close) any other point on the surface. On the pattern hull, the notion of gauge invariance for quasiperiodic eigenfunctions mentioned above is similar to gauge invariance in a Brillouin zone, taking $\ket{k}\rightarrow\ket{k+\delta}$.

\indent Thus, we consider the space of continuous functions on the hull of the pattern $\mathcal{C}(\Omega)$, the direct analog of a Brillouin zone, and introduce dynamics by defining the action of pattern translations, $\tau$, on these functions, defining a so-called $C^{*}$-Algebra, $\mathcal{C}(\Omega)\rtimes_{\tau}\mathbb{G}$, where $\mathbb{G}$ is the group generated by translations. Elements of this unital algebra are the non-commutative analogs for Hamiltonians on the momentum space torus. Adding information about the on-site Hilbert space - bands in translation invariant case - and a differential operator extracts sufficient information from the quasiperiodic system to define topology in the same way as done in conventional translation invariant band topology. More specifically, one relies on a generalization of the Atiyah-Singer Index theorem to spectral triples by Connes and Teleman \cite{connes1994quasiconformal}. 

\indent As discussed in S.I.\ref{siaahalgebra}, the unital algebra generated by a quasiperiodic pattern with real-space translations is a non-commutative n-torus -- dimension corresponding to the number of generators for the pattern and system dynamics. As a direct consequence, the spectral gaps of quasiperiodic systems as a function of the incommensurate parameter, $\Theta \in [0,1]$, can be labeled by integers, see Fig.~\ref{patternfigure}a, i.e. $\lbrace m+n\Theta\vert m,n\in\mathbb{Z}\rbrace\cap[0,N]$ with N is system size, label the gaps in the IDoS of the AAH Model. By construction, this labeling is invariant to small disorder as long as the pattern is well defined and gaps remain open \cite{bourne2018non,prodan2015,prodan2013non}.

\section{S.I. Generalized Andre-Aubry-Harper Model}\label{gaahexamplesi} 
While the AAH model is the canonical example of a 1D quasi-periodic systems, there are many generalizations of the Hamiltonian, modifying the on-site potential into other quasi-periodic patterns and/or dressing the hopping terms. These generalizations introduce new features to both the spectrum and wave-functions. Here we focus on a particular generalization parameterized by the onsite potential,
\begin{eqnarray}
V(x) = 2V \frac{\cos(\Theta x+\delta_{x})}{1-b\cos(\Theta x+\delta_{x})},
\end{eqnarray}
where $b\in(-1,1)$ detunes the model from the AAH model. As mentioned in the main text, this model hosts a \textit{mobility edge} -- states undergo a localized to delocalized transition at a fixed energy. It originates from its modified duality \cite{ganeshan2015nearest}, which depends on the energy $E$ of the eigenfunction in question
\begin{eqnarray}
b E = \textnormal{2 sgn}(\lambda)(\vert t\vert-\vert\lambda\vert)
\end{eqnarray}
Here, we reproduce this duality transformation from \cite{ganeshan2015nearest} below to show the immediate benefits from the above perspectives. In particular, the duality does not survive for Liouville $\alpha$, and the mobility edge arise from the absence of a gauge transformation reflecting the 1D duality in the 2D parent Hamiltonian. 

We begin with the full Hamiltonian,
\begin{eqnarray}\label{gaahsi}
\hat{H} = \sum_{x}t\hat{c}^{\dagger}_{x+1}\hat{c}_{x} +t^{*}\hat{c}^{\dagger}_{x}\hat{c}_{x+1} + 2V \frac{\cos(\Theta x+\delta_{x})}{1-b\cos(\Theta x+\delta_{x})}.\nonumber\\
\end{eqnarray}
Then, and rewrite it in the following way acting on a particle at site $x$,
\begin{eqnarray}\label{gaahduality}
tu_{x+1}+t^{*}u_{x-1} + g\chi_{x}(\beta,\delta_{x})u_{x} = (E + 2V\cosh{\beta})u_{x}\nonumber\\
\end{eqnarray}
with $\cosh{\beta} = 1/b$, $g = 2V\cosh^{2}{\beta}/\sinh{\beta}$, and 
\begin{eqnarray}\label{onsitepotenstialsi}
\chi_{x}(\beta,\delta_{x}) = \frac{\sinh{\beta}}{\cosh{\beta}-\cos{(\Theta x+\delta_{x})}}
\end{eqnarray}
Note, all sign dependence can be modulated by choosing a phase shift $\delta_{x}$. While \cite{ganeshan2015nearest} uses this to simplify the duality transformation, we notice this implies a phase shift changes the duality condition by changing the relative signs of $\cosh{\beta}$ and V. Continuing, one can decompose the onsite, potential further
\begin{eqnarray}\label{geomsumsi}
\chi_{x}(\beta,\delta_{x}) = \sum_{r=-\infty}^{\infty}e^{-\beta\vert r\vert}e^{ir(\Theta x+\delta_{x})}
\end{eqnarray}
In this form, it becomes clear that a Fourier-like transform will result in a new effective hopping term.
We first apply the transformation 
\begin{eqnarray}\label{GAAHtransformsi1}
b_{n} = \sum_{x}e^{i(\Theta nx)}u_{x}
\end{eqnarray}
\begin{widetext}
To keep the phase dependences, we choose $t = \tau e^{i\delta_{k}}$. The resulting Hamiltonian is:
\begin{eqnarray}\label{gaahduality1}
2\tau\cos{(\Theta n+\delta_{k})}b_{n} + \sum_{x}ge^{i\Theta nx}\chi_{x}(\beta)u_{x} &=& \sum_{x}(E + 2V\cosh{\beta})e^{i(\Theta nx)}u_{n}\nonumber\\
\sum_{x}ge^{i\Theta nx}\sum_{r=-\infty}^{\infty}e^{-\beta\vert r\vert}e^{ir(\Theta x+\delta_{x})}u_{x} &=& (E+2V\cosh{\beta} - 2\tau\cos{(\Theta n+\delta_{k})})b_{n}\nonumber\\
g\sum_{r=-\infty}^{\infty}e^{-\beta\vert r-n\vert}e^{i(r-n)\delta_{x}}b_{r} &=& (E+2V\cosh{\beta} - 2\tau\cos{(\Theta n+\delta_{k})})b_{n}\nonumber\\
g\sum_{r=-\infty}^{\infty}e^{-\beta\vert r-n\vert}e^{i(r-n)\delta_{x}}b_{r} &=& \omega\chi^{-1}_{n}(\beta_{0},\delta_{k})b_{n}
\end{eqnarray}
Where, $\omega = 2\tau\sinh{\beta_{0}}$ and $\cosh{\beta_{0}} = 1/2\tau(2V\cosh{\beta}+E)$. One can then apply the transformation:
\begin{eqnarray}\label{GAAHtransformsi2}
v_{m} = \sum_{n}e^{i\Theta mn}\chi^{-1}_{n}(\beta_{0},\delta_{k})b_{n}
\end{eqnarray}
Resulting in 
\begin{align}\label{gaahduality2}
\sum_{n}g\sum_{r'=-\infty}^{\infty}e^{-\beta\vert r'-n\vert}e^{i(r'-n)\delta_{x}}e^{i\Theta mn}b_{r'} = \sum_{n}\omega\chi^{-1}_{n}(\beta_{0},\delta_{k})e^{i\Theta mn}b_{n}\nonumber\\
\sum_{n}g\sum_{r'=-\infty}^{\infty}e^{-\beta\vert r'-n\vert}e^{i(r'-n)\delta_{x}}e^{i\Theta m(n-r')}\sum_{r=-\infty}^{\infty} e^{-\beta_{0}\vert r\vert}e^{ir(\Theta r'+\delta_{k})} e^{i\Theta mr'}\chi_{r'}^{-1}(\beta_{0},\delta_{k})b_{r'} = \omega v_{m}\nonumber\\
g\sum_{r=-\infty}^{\infty} e^{-\beta_{0}\vert r-m\vert}e^{i(r-m)\delta_{k}} v_{r} = \omega \chi^{-1}_{m}(\beta,-\delta_{x})v_{m}
\end{align}
We can now take the final step and define $f_{k} = \sum_{m}(e^{i\Theta mk}) v_{m}$, allowing us to rewrite the Hamiltonian as:
\begin{eqnarray}\label{gaahduality3}
\sum_{m}g\sum_{r=-\infty}^{\infty} e^{-\beta_{0}\vert r-m\vert}e^{i(r-m)\delta_{k}} e^{i\Theta mk} v_{r} &=& \omega \sum_{m}\chi^{-1}_{m}(\beta,-\delta_{x})e^{i\Theta mk}v_{m}\nonumber\\
\sum_{r}g\sum_{m=-\infty}^{\infty} e^{-\beta_{0}\vert r-m\vert}e^{i(r-m)\delta_{k}} e^{i\Theta (m-r)k}e^{i\Theta kr} v_{r} &=& \omega \sum_{m}\chi^{-1}_{m}(\beta,-\delta_{x})e^{i\Theta mk}v_{m}\nonumber\\
g\chi_{k}(\beta_{0},-\delta_{k}) \sum_{r}e^{i\Theta kr} v_{r} &=& 2\tau\sinh{\beta_{0}} \sum_{m}\frac{\cosh{\beta}-\cos{(\Theta m+\delta_{x})}}{\sinh{\beta}}e^{i\Theta mk}v_{m}\nonumber\\
(\tau e^{i\delta_{x}}f_{k+1}+\tau e^{-i\delta_{x}}f_{k-1})+g\frac{\sinh{\beta}}{\sinh{\beta_{0}}}\chi_{k}(\beta_{0},-\delta_{k}) f_{k} &=& 2\tau\cosh{\beta}f_{k}
\end{eqnarray}
Thus, as shown in \cite{ganeshan2015nearest}, if $\beta_{0} = \beta$ (and $\delta_{k}=-\delta_{x}$), the new Hamiltonian under the complete transformation,
\begin{eqnarray}\label{dualitytransformationsi}
f_{k} = \sum_{m,n,x} e^{i\Theta(mk+mn+nx)}\chi_{n}^{-1}(\beta_{0},\delta_{k})u_{x},
\end{eqnarray} 
is self dual. This results in the cited condition,
\begin{eqnarray}\label{derivedcondition}
2V/b + E = 2\tau/b \implies bE = 2(\tau-V).
\end{eqnarray}
This result is an exciting example of a quasi-periodic system with a mobility edge, and suggestive of a connection between such models and random disorder. However, examining the problem from the pGF perspective presented here, we see the constraints are indeed topological unlike random disorder. The proof follows as in section S.I.~\ref{siboundstransfer}.
\begin{figure*}
    \centering
    \includegraphics[scale = .5]{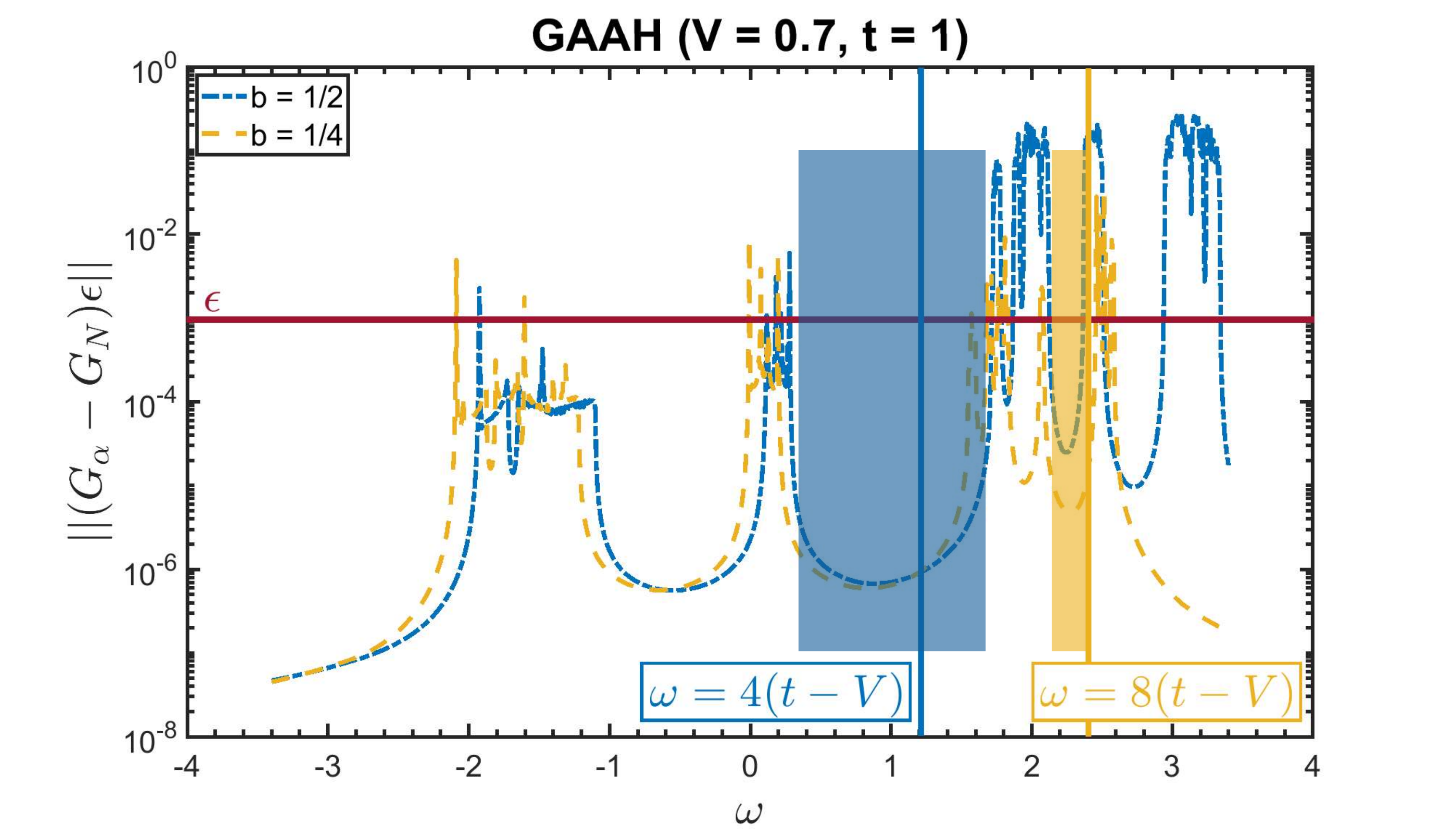}
    \caption{Plot of pGF convergence for GAAH model at $V = 0.7, t = 1$ for $b = 1/2, 1/4$ and $N = 4096$. We notice the shift in mobility edge from convergence criteria. Note, blue/yellow lines indicate analytic mobility edge, and blue/yellow boxes indicate spectral gaps in which those mobility edges line. The blue line occurs deep in a spectral gap for which states above are localized and states below are delocalized, while the yellow line is closer to the gap edge. Since the the quasi-periodic spectrum forms a cantor set, all mobility edges will fall in a spectral gap, but numerical precision of pGF convergence is better in large gaps}
    \label{fig:gaah}
\end{figure*}
\subsection{GAAH Projected Green's Function}
We begin by constructing the 2D rational approximates. In Eq.~\eqref{GAAHduality} we can redefine $E' = E + 2V\cosh{\beta}$ and operators can be assigned a corresponding phase, $\delta_x$. Then, using the expansion in Eq.~\eqref{geomsumsi}, we perform an inverse Fourier transform to arrive at 

\begin{eqnarray}\label{2DlongrangeGAAHsi}
    \mathcal{H}_{2D} &=& \sum_{x,\delta_x} \left[ t\hat{c}_{x+1,\delta_x}^{\dagger}\hat{c}_{x,\delta_x} +  h.c.+g \sum_{r\geq0}e^{-\beta\vert r\vert} \cos{(\Theta x r +\delta_x)}\hat{c}^{\dagger}_{x,\delta_x}\hat{c}_{x,\delta_x} \right]\nonumber\\
    \mathcal{H}_{2D} &=&\sum_{x,y} \left[ t\hat{c}_{x+1,y}^{\dagger}\hat{c}_{x,y} + g/2 \sum_{r = -\infty}^{\infty}e^{-\beta\vert r\vert} e^{ir\Theta x}\hat{c}^{\dagger}_{x,y+r}\hat{c}_{x,y} + h.c.\right]
\end{eqnarray}

Notice, that the 2D parent Hamiltonian of the GAAH model has long-range hopping along the ``phase" coordinate and short range hopping along the ``real space" coordinate. As such, no simple gauge transformation will generate the 1D duality transformation in 2D. And, the rational approximates from S.I.~\ref{siboundstransfer} will only host finite unit cells for a horizontal unit cell choice.

We take the approximating sequence,
\begin{eqnarray}\label{2DlongrangeGAAHrationalssi}
    \mathcal{H}_{2D} &=& \sum_{x,\delta_x} \left[ t\hat{c}_{x+1,\delta_x}^{\dagger}\hat{c}_{x,\delta_x} +  h.c.+g \sum_{r\geq0}e^{-\beta\vert r\vert} \cos{(\frac{p_N}{q_N} x r +\delta_x)}\hat{c}^{\dagger}_{x,\delta_x}\hat{c}_{x,\delta_x} \right],
\end{eqnarray}
where $\frac{p_N}{q_N}$ is the $N$-th continued fraction approximation of $\alpha$. We bound the difference in onsite potentials for the rational approximates vs $\mathcal{H}_{2D}$ as in S.I.~\ref{siboundstransfer}, using
\begin{eqnarray}
\vert2g\cos{(2\pi\alpha x r +\delta_x)} -2g\cos{(2\pi\alpha x r +\delta_x)}\vert < \vert 4g \sin{(\pi\delta_N rx)} < \vert 4g\pi\delta_N q_N r\vert,
\end{eqnarray}
where $\delta_N = \vert\alpha -\frac{p_N}{q_N}\vert$ and for $\alpha$ diophantine $\delta_N<1/\sqrt{5}q_{N}$. Unlike above, the dependence on $r$ makes this term unbounded, but we have the exponential term, $e^{-\beta\vert r\vert}$, which reduces the difference to
\begin{eqnarray}
\vert2g\chi_{\alpha}(\beta,\delta_x) - \chi_{N}(\beta,\delta_x)\vert <  \vert\sum_{r\geq 0} 4g\pi\delta_N q_N r e^{-\beta\vert r\vert} \vert = \vert 4g\pi\delta_N q_N (\partial_\beta e^{-\beta r})\vert = \vert 4g\pi\delta_N q_N \vert\frac{e^{-\beta}}{(1-e^{-\beta})^{2}}< \frac{C}{q_N}.
\end{eqnarray}
The same argument regarding an $i\epsilon$ prescription applies and we need to bound 
\begin{eqnarray}
\vert\vert(\omega-\mathcal{H}_{2D,N})\vert\vert< C\sqrt{q_N}
\end{eqnarray}
Unlike above, where self-duality meant we had not hope of a stronger bound, here we can keep $\omega$ to help us get a tighter bound on the convergent parameter space. Notice the arithmetic mean of $\cosh{\beta}\vert\chi(\beta)/\sinh{\beta}\vert$ over all $x$ is just the inverse geometric mean of $\vert 1 -b\cos(\Theta x)\vert$. So, $\det{\vert\mathcal{H}_{2D,N}\vert}< [(2(V-t)/b))^{q_{N}}$
We obtain:
$\det(\omega - \mathcal{H}_{2D,N}) <\infty$ if 
$$\omega + 2V/b  < 2t/b$$
For horizontal unit cells and $\alpha$ diophantine, the pGF converges when $bE<2(t-V)$ with positive $t,V$, see Fig.~\ref{fig:gaah}.

\end{widetext}

\end{document}